\def\spose#1{\hbox to 0pt{#1\hss}}

\def\multleft#1{\hbox to size{\vbox {\halign {\lft{##}\cr #1}}\hfill}\par}
\def\multright#1{\hbox to size{\vbox {\halign {\rt{##}\cr #1}}\hfill}\par}

\def\today{\ifcase\month\or January\or February\or March\or April\or May\or
      June\or July\or August\or September\or October\or November\or December\fi
      \space\number\day, \number\year}

\def\H2{\hbox{H$_{2}$}}

\documentclass{mn2e}
\usepackage{times}
\usepackage{amssymb}
\usepackage{epsfig}
\usepackage{lscape}
\usepackage{graphicx}
\begin{document}
\hsize=6truein
          
\title[The $\bmath{z=9}$ galaxy population]
{The $\bmath{z=9-10}$ galaxy population in the Hubble Frontier Fields and CLASH surveys: The $\bmath{z=9}$ LF and further evidence for a smooth decline in UV luminosity density at $\bmath{z\geq8}$}
\author[D.J.~McLeod et al.]
{D. J. McLeod$^{1}$\thanks{Email: mcleod@roe.ac.uk}, R. J. McLure$^{1}$,
J. S. Dunlop$^{1}$
\footnotesize\\
$^{1}$SUPA\thanks{Scottish Universities Physics Alliance}, Institute
for Astronomy, University of Edinburgh, Royal Observatory, Edinburgh
EH9 3HJ}
\maketitle

\begin{abstract}
We present the results of a search for $z=9-10$ galaxies within the first 
8 pointings of the Hubble Frontier Fields (HFF) survey (4 clusters plus 4 
parallel fields) and 20 cluster fields from the CLASH survey. Combined 
with our previous analysis of the Hubble Ultra-Deep field (HUDF), we have 
now completed a search for $z=9-10$ galaxies over 
$\simeq 130$\,arcmin$^2$, spread across 29 {\it HST} WFC3/IR pointings. As 
in our recent study of the first two HFF fields, 
we confine our primary search for high-redshift candidates in the 
HFF imaging to the uniformly deep (i.e. $\sigma_{160}>30$ AB mag in 
0.5-arcsec diameter apertures), relatively low magnification regions. In 
the CLASH fields our search was confined to uniformly deep regions where 
$\sigma_{160}>28.8$ AB mag. Our spectral energy 
distribution fitting analysis unveils a sample of 33 galaxy candidates at $z_{phot}\geq 
8.4$, five of which have primary photometric 
redshift solutions in the range $9.6<z_{phot}<11.2$. By calculating a 
de-lensed effective volume for each candidate, the improved statistics and 
reduced cosmic variance provided by our new sample allows a more 
accurate determination of the UV-selected galaxy luminosity function (LF) 
at $z\simeq 9$. Our new results strengthen our previous conclusion that 
the LF appears to evolve smoothly from $z=8$ to $z=9$, an evolution which 
can be equally well modelled by a factor of $\simeq 2$ drop in density, or 
a dimming of $\simeq 0.5$ mag in $M^{\star}$. Moreover, based on 
our new sample, we are able to place initial constraints on the $z=10$ LF, 
finding that the number density at $M_{1500}\simeq-19.7$ is $\log(\phi) = 
-4.1^{+0.2}_{-0.3}$, a factor of $\simeq 2$ lower than at $z=9$. 
Finally, we use our new results to re-visit 
the issue of the decline in UV luminosity density ($\rho_{UV}$) at $z\geq 
8$. We conclude that the data continue to support a smooth decline in 
$\rho_{UV}$ over the redshift interval $6<z<10$, in agreement 
with simple models of early galaxy evolution driven by the growth 
in the underlying dark matter halo mass function.
\end{abstract}


\section{INTRODUCTION}
Our understanding of cosmic history within the first billion years after the Big Bang 
has improved dramatically in recent times, largely thanks to the near-infrared capabilities of the \textit{Hubble Space Telescope (HST)} 
as realised through Wide Field Camera 3 (WFC3/IR).  It has now become
relatively routine to assemble catalogues of galaxies out to $z \simeq
8$, either via colour-colour selection or spectral energy distribution
(SED) fitting (e.g. McLure et al. 2010, 2011; Bouwens et al. 2011;
review by Dunlop 2013),  with the current frontier of observations now at $z\sim9-10$ (e.g. McLure et al. 2013; Oesch et al. 2014; McLeod et al. 2015; Bouwens et al. 2015).

An important measure of the evolution of the high-redshift galaxy population 
is the UV-selected galaxy luminosity function (LF).  To determine the form and evolution of the UV galaxy LF, it is crucial 
to establish a wide baseline in UV luminosity,  and this can now be achieved by combining the ultra-deep imaging 
provided by the Hubble Ultra Deep Field (HUDF) with the shallower, wider data of the CANDELS fields, and the still wider, shallower ground-based 
imaging obtained in the UKIDSS Ultra Deep Survey (UDS) and UltraVista/COSMOS fields (McCracken et al. 2012; Bowler et al. 2012, 2014, 2015).  
Other surveys, such as the pure-parallel {\it HST} BoRG survey (Trenti et al. 2011) also provide the opportunity to study the 
bright end of the LF at high redshift via multiple, shallow, pencil-beam pointings.  The LF at redshift $z=6-8$ has been the 
subject of numerous studies in the recent literature, and is now largely well-understood at magnitudes fainter than $M_{1500}\simeq-21$ 
(see McLure et al. 2013; Finkelstein et al. 2014; Bowler et al. 2014, 2015; Bouwens et al. 2015).

By integrating the luminosity-weighted LF, one arrives at the UV luminosity density ($\rho_{UV}$), which can then be converted into a 
star-formation rate density ($\rho_{SFR}$; see Kennicutt \& Evans 2012).  As expected, given the agreement in the form of the faint end of the LF, 
the determinations of $\rho_{UV}$ are in general concordance out to $z\sim8$.  Due to observational constraints, 
the determination of these quantities at even earlier epochs is challenging, and different studies have reached different conclusions.

The first meaningful study of the $z \simeq 9$ galaxy population was undertaken 
by Ellis et al. (2013) and McLure et al. (2013), based on the small sample of $z \simeq 9$ galaxies uncovered within the HUDF12 dataset
(which featured the key ultra-deep $Y_{105}, J_{125}, J_{140}\,\&\, H_{160}$ imaging necessary to identify robust $z \simeq 9$ 
galaxy candidates; Koekemoer et al. 2013).  Based on this sample,
Ellis et al. (2013) concluded that the decline in $\rho_{UV}$ beyond
$z\simeq4$ continued relatively smoothly beyond $z \simeq 8$,
indicative of a gentle decline of star-formation rate density back to earlier
epochs, as naturally required to consistently explain cosmic reionization 
and recent constraints on Thomson scattering optical
depth delivered by WMAP and Planck (e.g. Robertson et al. 2013,
2015). However, in contrast, Oesch et al. (2014) conclude that $z \simeq 8$ marks the onset of a 
dramatic fall-off in $\rho_{UV}$, at which the decline of luminosity
density switches suddenly from $\rho_{UV} \propto (1+z)^{-3.6}$, 
to a much more rapid $\rho_{UV} \propto(1+z)^{-10.9}$.

To resolve this disagreement, what is required is more deep near-infrared imaging, including 
the crucial $J_{140}$ WFC3/IR filter, and ideally involving many different sightlines to minimize 
the uncertainty introduced by cosmic variance. Such imaging is now being 
delivered by the Hubble Frontier Fields (HFF) programme, from which the {\it HST} WFC3/IR 
imaging has now been released for four clusters and their corresponding parallel fields. 
The HFF programme has already proven to be instrumental in furthering our knowledge of 
galaxies in the young Universe, with a number of studies already published on high-redshift 
galaxy populations (e.g. Zheng et al. 2014; Zitrin et al. 2014; Atek et al. 2015; 
Ishigaki et al. 2015).  One of the primary aims of the HFF survey is
to exploit the gravitational lensing provided by the cluster fields to
aid the discovery and study of faint galaxies at high redshift. However, as discussed in McLeod et al. (2015) and further below, 
the parallel pointings are arguably more powerful than the cluster pointings for advancing 
our knowledge of the galaxy UV LF at extreme redshifts.

In McLeod et al. (2015) we analysed the imaging from the first two HFF cluster+parallel pointings to derive 
improved constraints on the $z \simeq 9$ LF and revisit the issue of
the decline in $\rho_{UV}$ at $z\geq8$. We uncovered twelve galaxies in the redshift range $8.4<z<9.5$ from within 
the area covered by these four WFC3/IR pointings, enabling us to place the 
best constraints to date on the evolution of the LF from $z \simeq 8$ to $z \simeq 9$. 
Our main finding was that the decline in $\rho_{UV}$ beyond $z \simeq 8$ was smooth and relatively modest,
with  $\rho_{UV}$ falling off much less steeply than $\rho_{UV} \simeq (1+z)^{-10.9}$ as claimed by Oesch et al. (2014). 
Indeed we found that the decline of $\log_{10} \rho_{UV}$ from $z \simeq 6$ to $z \simeq 9$ could be well described 
by a simple linear function of redshift, extrapolation of which to $z \simeq 10$ agreed with the (albeit highly uncertain) 
$z \simeq 10$ determination of Oesch et al. (2014).

However, we are still in the regime of small number statistics at $z \simeq 9 - 10$, and more data are still required 
to resolve this important issue. We thus here present a new study of galaxy evolution beyond $z \simeq 8$ which 
improves on that presented in McLeod et al. (2015) in several respects. Firstly, the available HFF dataset has now been doubled 
by the release of the imaging of the third and fourth cluster and parallel fields (i.e. of  MACS0717 and MACS1149), 
and we have now extended our search for $z \simeq 9$ galaxies to include these new WFC3/IR pointings. 
Secondly, while our most robust results transpire to be driven primarily
by the homogeneous survey area provided by the HFF parallel fields, we have also undertaken a thorough 
analysis of more strongly lensed regions of the imaging, and in particular have utilised this approach to search for 
$z \simeq 9$ galaxies in twenty fields from the CLASH survey. Finally, we have now also searched both the HFF and CLASH datasets for galaxies at $z \simeq 10$. 

As important as lensed fields are for searching for the earliest galaxies, especially intrinsically faint objects, 
extracting robust quantitative information on high-redshift galaxy evolution from gravitationally 
lensed fields is challenging and often problematic.  A number of groups 
have provided lensing maps for each of the HFF clusters, and similarly Zitrin et al. (2013) 
have provided two alternative lensing maps for all of the CLASH fields.  Unfortunately, the magnifications predicted 
by the different models are often significantly different, and these discrepancies are often particularly severe 
in the high magnification regions around the critical lines.  This does not just affect the individual candidates 
in a sample, but also the effective selection volume that one derives for a particular lensed field.  
Due to these effects, a determination of the LF from a strongly lensed field can be drastically wrong 
unless the lensing uncertainties can be controlled. Finally, the intracluster light often 
significantly reduces the effective photometric depth for much of the on-cluster image unless it can be reliably removed.

Due to these issues, in McLeod et al. (2015) we decided to focus our attention on the 
clean, homogeneously deep regions of the imaging in the 
Abell2744 and MACS0416 fields.  While this approach reduced the available search 
area somewhat, it crucially allowed us to derive well-defined samples from robust cosmological volumes determined
from regions of the imaging with negligible or low/well-constrained
magnification factors ($\mu$). In the new study presented here, we have retained this deliberately conservative approach to derive our most robust new 
determination of the $z \simeq 9$ LF, now based on applying this method to all eight available HFF pointings.

However, in an attempt to exploit all the available deep $J_{140}$
imaging, in this study we have supplemented the eight HFF pointings with a search for $z \simeq 9 - 10$ galaxies in the 
CLASH survey (Postman et al. 2012).  Although the WFC3/IR imaging of the CLASH clusters 
is significantly shallower than the HFF imaging, the effective depth of much of the lensed CLASH imaging is comparable 
to the depth of the unlensed imaging in the HFF parallel fields. Notwithstanding the aforementioned
concerns about uncertain lensing magnifications, an attraction of incorporating the CLASH imaging 
is that it provides a large number of widely separated pointings on the sky, potentially reducing the impact of cosmic 
variance (see Trenti \& Stiavelli 2008).

\begin{figure*}
\includegraphics[width=18.1cm, angle=0]{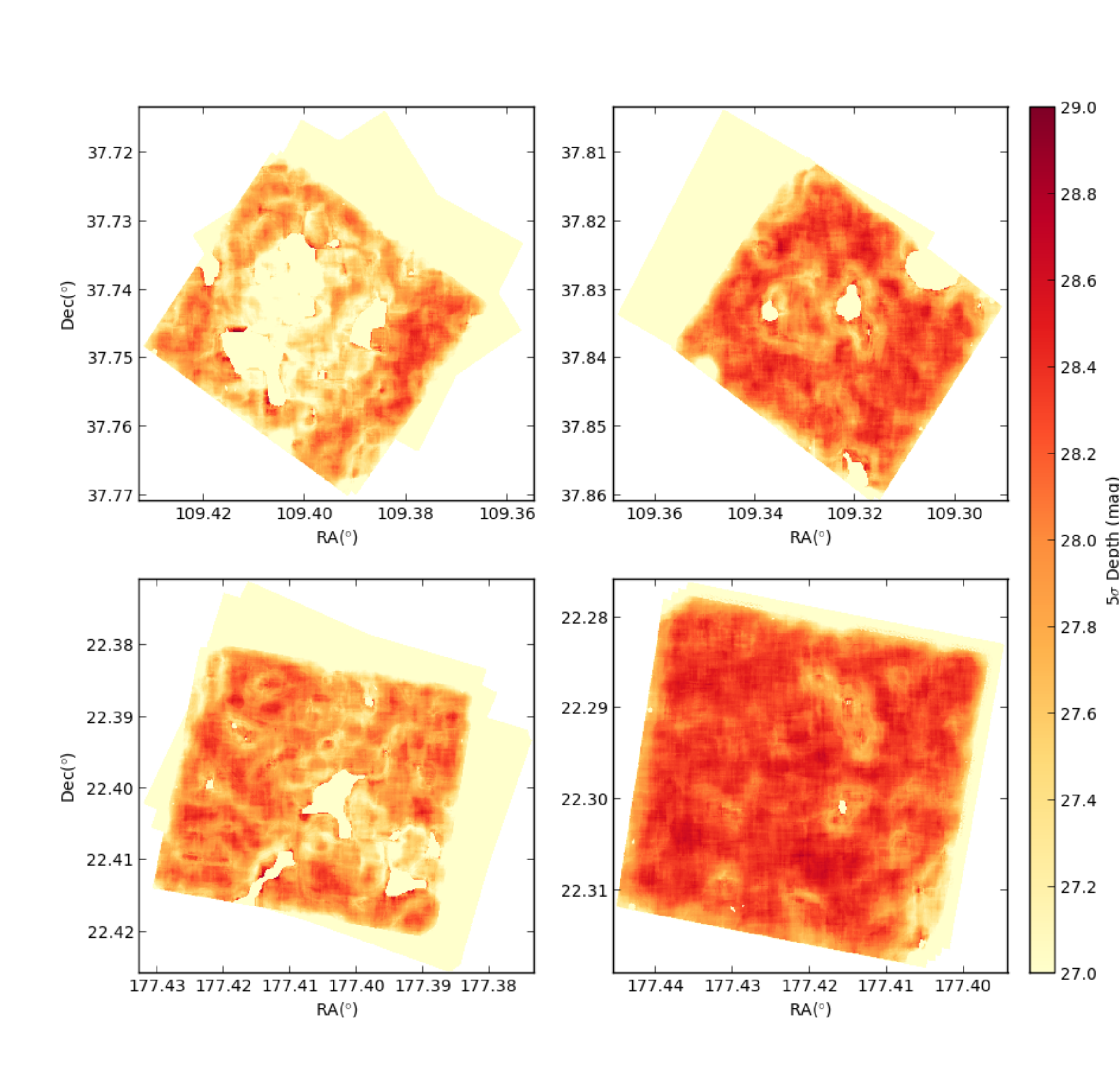}
\caption{Colour coded $H_{160}$ local depth maps for each of the four new HFF pointings; i.e. 
the MACS0717 (top) and MACS1149 (bottom) cluster fields and their associated parallel fields.  
Local depths are 5$\sigma$ measurements for 85\% of total point-source flux, based on the rms from a 
grid of 150-200 apertures in the vicinity of each object.  For the blank-field methodology used in this study, 
cuts to the survey area were implemented, culling regions where the
$5\sigma$ depth was not at least $H_{160}= 28$\,mag in this format.  For the cluster fields, the effective area is further reduced 
by the effect of the gravitational lensing of the foreground cluster.  Notably, the MACS0717 
cluster field has severe crowding and so is generally shallower than the rest of the cluster fields, 
providing little extra accessible cosmic volume. The shallower imaging around the edges, due to dithering, 
was excluded from the search and has been set to a magnitude floor of 27.0 for display purposes. Comparable 
plots for the first four HFF pointings can be found in McLeod et al. (2015).}
\end{figure*}

This paper is structured as follows. In Section 2 we discuss the HFF
and CLASH imaging data, and provide a brief description of the
selection of our high-redshift galaxy candidates. This includes a discussion of 
how we have used deconfused longer-wavelength photometry from 
\textit{Spitzer} IRAC imaging 
to aid in cleaning our sample of potential low-redshift 
contaminants. In Section 3, we present and discuss our final sample of $z\geq9$ galaxies and summarise their basic 
properties. In Section 4 we present our new, updated determination of the $z=9$ LF from the blank-field method, 
(i.e. constraining galaxy selection to the cleanest, homogeneous regions of the imaging). 
In Section 5 we extend our analysis to derive 
the lensed-field LF determination at $z \simeq 9$, 
and to provide new constraints on the LF at $z \simeq 10$.  
Finally, in Section 6 we present our new and 
improved measurement of the early rise of $\rho_{UV}$, and compare our results to a 
range of theoretical predictions. Our conclusions are summarised 
in Section 7.

Throughout the paper we will refer to the following {\it HST} ACS+WFC3/IR filters: F435W, F475W, F555W, F606W, F625W, F775W, F814W, 
F850LP, F105W, F110W, F125W, F140W \& F160W as $B_{435}$, $B_{475}$, $V_{555}$, $V_{606}$, $V_{625}$, $i_{775}$, $i_{814}$, 
$z_{850}$, $Y_{105}$, $J_{110}$, $J_{125}$, $J_{140}$ \& $H_{160}$ respectively. All magnitudes are quoted 
in the AB system (Oke 1974; Oke \& Gunn 1983).  For cosmological calculations, we adopt 
$\Omega_{0}=0.3, \Omega_{\Lambda}=0.7$ and $H_{0}=70$ kms$^{-1}$Mpc$^{-1}$.

\section{Data and Candidate Selection}
\subsection{Observations}
\subsubsection{Hubble Frontier Fields}

In this section we review the optical, near-IR and mid-IR data used in this study.  
From the HFF survey we utilised the first four clusters and their corresponding parallel fields.  
For each of these, {\it HST} imaging was carried out in three ACS wavebands ($B_{435}$, $V_{606}$, $i_{814}$) 
and four WFC3/IR wavebands ($Y_{105}$, $J_{125}$, $J_{140}$, $H_{160}$).  Global depths for 
each of the images were calculated from apertures corrected to enclose 85\% of the total 
flux, assuming a point source.  In the parallel fields, the median $5\sigma$ depths in this 
format were $B_{435}$=28.8, $V_{606}$=28.8, $i_{814}$=28.9, $Y_{105}$=28.9, $J_{125}$=28.4, $J_{140}$=28.3 
and $H_{160}$=28.2.  Note that the MACS0717 parallel field was typically 
$\sim0.1-0.3$ magnitudes shallower than these values, due to a higher background level.  
The HFF cluster fields were also typically shallower than these depths due to source crowding 
and intracluster light.  In the `clean' regions of the cluster imaging the depths are similar to 
those of the parallel fields.

For the first two cluster/parallel fields, Abell2744 and MACS0416, HAWK-I $K_{S}$ data at $2.2\mu$m 
(PI Brammer, ESO Programme ID 092.A-0472) was available.  For a 1$^{\prime\prime}$ diameter aperture, 
enclosing 75\% of total flux, the depth in $K_{S}$ was typically $\simeq$25.6 mag.  All eight HFF fields had 
\textit{Spitzer Space Telescope} Infrared Array Camera (IRAC) imaging (PI Soifer\footnote{Program IDs: 90257, 
90258, 90259 and 90260.}) available in channel 1 ($3.6\mu$m) and
channel 2 ($4.5\mu$m).  This data typically reaches a 
depth of $\simeq$24.5 mag (at 85\% of total), although this is specific to the cleaner parallel 
regions and is subject to the ability to perform effective deconfusion (see Section 2.4).

\subsubsection{CLASH}
From the CLASH survey we analysed 23 of the 25 available fields, omitting those which did not 
have any archival IRAC imaging (Abell1423 and CLJ1226) to avoid having any candidates for 
which there was no available long-wavelength photometry to check for low-redshift contamination. 
No public $K_{s}$-band imaging was available for the CLASH fields. Three of the CLASH fields are 
also HFF fields, and so our final sample only exploits the superior HFF data for these clusters.  
The CLASH imaging of these clusters was however still analysed in the same manner, and used 
in order to check for consistency.  Our final analysis of the CLASH
candidates is therefore confined to the twenty fields unique to the CLASH survey and with IRAC imaging.

The CLASH imaging used in this study consisted primarily of the optical bands $B_{435}$, $B_{475}$, 
$V_{606}$, $i_{775}$, $i_{814}$, $z_{850}$ and near-infrared wavebands $Y_{105}$, $J_{110}$, $J_{125}$, 
$J_{140}$ and $H_{160}$, although some CLASH fields have also been imaged in $V_{555}$ and $V_{625}$.  
The median 5$\sigma$ global depths (at 85\% of total assuming a point source) across the CLASH 
fields were found to be $B_{435}$=26.8, $B_{475}$=27.1, $V_{555}$=27.5, $V_{606}$=27.4, $V_{625}$=26.8, $i_{775}$=26.6, 
$i_{814}$=27.1, $z_{850}$=26.2, $Y_{105}$=26.8, $J_{110}$=26.9, $J_{125}$=26.7, $J_{140}$=26.7 and $H_{160}$=26.5, 
although in practice there are again significant variations in depth due to varying levels of 
source crowding and intracluster light.  The IRAC
imaging\footnote{Program IDs: 90009, 80168, 40652, 
60034, 545 and 50393.} used was in the $3.6\mu$m and $4.5\mu$m filters, with a 5$\sigma$ median depth for 
both channels of 23.9 mag (at 85\% point-source flux).  However, the range of $5\sigma$ depths 
varied between 22.5 and 24.2 mag.  The CLASH survey also features
WFC3/UVIS imaging in the F225W, F275W, F336W and F390W bands. This data was not included in our SED fitting analysis because the UVIS imaging is significantly shallower than the ACS data. 
Nevertheless, in compiling the final candidate list, the UVIS imaging was inspected as an extra check 
against possible contamination by low-redshift interlopers in the sample.  

In total, the raw search area provided by the eight HFF pointings, the HUDF and the twenty CLASH pointings 
totals $\simeq$ 130 arcmin$^{2}$.

\subsection{Photometric catalogues}
For each of the survey fields, initial photometric catalogues were constructed by running 
{\sc SExtractor} 2.8.6 (Bertin \& Arnouts 1996) in dual-image mode, using each of the 
WFC3/IR images longward of the Lyman break at $z\geq8.5$ as detection images.  
Additionally, stacks of the relevant near-IR imaging ($J_{125}+J_{140}+H_{160}$, $J_{140}+H_{160}$ \& $J_{125}+J_{140}$) 
were also used as detection images, in order to boost the sensitivity to $z\geq9$ objects.

Every object detected at $\geq5\sigma$ in each catalogue was retained, and a master catalogue was 
produced containing every unique object from all of the detection catalogues.  For each object, 
the highest signal-to-noise detection was propagated to the final catalogue.  Given the position 
of the redshifted Lyman break, it is expected that no candidate would be detected in the optical bands, 
and so candidates were required to be undetected at the $\leq2\sigma$  level in every filter 
shortward of $Y_{105}$.  Various stacks of the optical photometry were also produced in order 
to further ensure that candidates were robustly undetected shortward of $Y_{105}$.

The photometry was measured in 0.4$^{\prime\prime}$ diameter apertures in the ACS and $Y_{105}$ filters 
and in 0.44$^{\prime\prime}$, 0.44$^{\prime\prime}$, 0.47$^{\prime\prime}$ and 0.5$^{\prime\prime}$ apertures 
for the $J_{110}$, $J_{125}$, $J_{140}$ and $H_{160}$ imaging respectively.  
This ensures $\geq$70\% of point-source flux is included in the near-IR bands and 
80-85\% of point-source flux in the ACS bands.  All fluxes were then corrected to a uniform 85\% of 
total point-source flux for the purposes of SED fitting.

\subsection{Local depth analysis}
Accurate photometric errors are crucial for obtaining a robust sample of high-redshift galaxies.  
Firstly, they are used in the optical flux cuts to remove possible low-redshift contaminants, 
and in the detection band magnitude cut to remove objects that are too faint.  
Secondly, the SED fitting requires robust photometric errors in order to produce accurate 
photometric redshifts.  Every lensed field in this study is subject to significant variations 
in the background and depth across the image, which is exacerbated by source crowding and intracluster light.  
Although less of an issue in the parallel fields, even here there are still slight variations in the depth.  
Following McLeod et al. (2015), local aperture-to-aperture rms depths were calculated for 
every object in the photometric catalogues, based on large grids of apertures placed 
on blank sky regions.  Based on the rms of the closest 150--200 apertures it was possible to 
calculate a robust local depth for every object, in every filter.  These aperture-to-aperture rms 
measurements are more accurate in capturing the various systematics than the drizzled rms maps 
supplied with the data.

By calculating the estimated local depth across the fields, detailed
depth maps were produced for each HFF and CLASH field.  The depth maps for the new MACS0717 and MACS1149 images 
and their respective parallel fields are shown in Fig.\,1, and
illustrate the reduced depth in the central regions of the galaxy clusters.  It is notable how severe this is for 
the MACS0717 cluster field, for which the WFC3/IR imaging is positioned such that the 
field-of-view is almost completely filled by the cluster.  Due to the
presence of a bright star and 
multiple large bright foreground galaxies, the MACS0717 parallel image also suffers a significant 
loss of useful survey area compared to the other HFF parallel images.

\subsection{Longer wavelength photometry}
In this study we include, where available, VLT HAWK-I $K_{S}$ and {\it Spitzer} IRAC $3.6\mu$m and $4.5\mu$m data.  
This is primarily to guard against low-redshift interlopers contaminating our $z\geq9$ galaxy sample.  
As discussed in McLeod et al. (2015), secondary photometric redshift solutions typically correspond 
to significantly reddened $z\simeq2$ galaxies, or galaxies at $z \simeq 6$ with quiescent stellar populations.  
With the added longer-wavelength photometry, these alternative redshift solutions can usually be 
robustly excluded or confirmed.  Even though the available imaging is significantly shallower 
than the shorter-wavelength {\it HST} imaging, non-detections can be vital in this process.

Obtaining accurate photometry in the IRAC bands is notoriously difficult given the (relatively) 
low spatial resolution and the resulting heavy blending of objects in
deep imaging.  In this study we therefore employed the deconfusion code \textsc{tphot} (Merlin et al. 2015) 
in order to extract IRAC and $K_{s}-$band photometry. In brief, this code utilizes the positional and surface-brightness information of
galaxies detected in a high spatial resolution image (HRI) to produce a
model of the corresponding low spatial resolution image (LRI). Two
dimensional galaxy templates are extracted from the HRI and convolved
with a transfer kernel to match the resolution of the LRI. These
low-resolution templates are then used to produce a best-fitting model
of the LRI image, in which the flux of each template is fit simultaneously. 
In this study we used the $J_{125}$+$J_{140}$+$H_{160}$ stack as the
HRI prior image because it provides the highest signal-to-noise ratio templates
for $z\simeq 9$ objects.  Full details of the \textsc{tphot} algorithm can be found in Merlin et al. (2015).

\begin{figure*}
\includegraphics{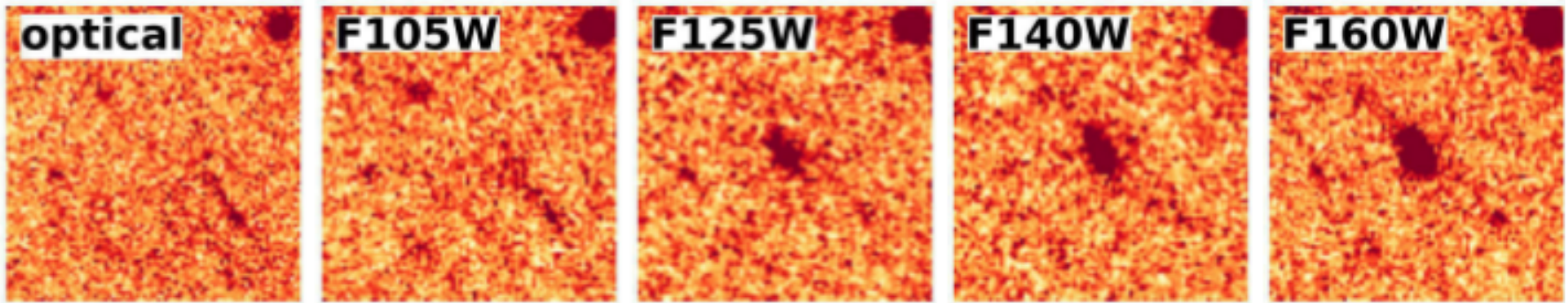}
\caption{{\it HST} imaging postage stamps (5$^{\prime\prime}$ x
  5$^{\prime\prime}$, orientation North=up and East=left) for
  HFF4C-9-1, a $z_{phot}=9.3$ candidate from the MACS1149 cluster field.  Left to right: optical stack, $Y_{105}$, $J_{125}$, $J_{140}$ and $H_{160}$ images.  
Both the stacked optical and $Y_{105}$ images are completely blank,
with the object securely detected in all three longer-wavelength images.
This constrains the break in the spectrum to be at $\lambda\simeq\rm1.25\mu$m (see Fig. 3).  The colour grayscale is set to the median background flux 
$\pm2\sigma$, once any flux from neighbouring objects has been excluded using the segmentation map. }
\end{figure*}

\subsection{Photometric redshifts}

Once the catalogue was culled of potential low-redshift objects with 
$\geq$2$\sigma$ optical detections, a photometric redshift analysis was performed.  
For this the SED template fitting code LePhare\footnote{www.cfht.hawaii.edu/~arnouts/LEPHARE/lephare.html} 
(Arnouts \& Ilbert 2011) was used.  The SED template 
library was based on the models of Bruzual \& Charlot 2003 (BC03), utilising IGM absorption 
according to the prescription by Madau (1995) and a Calzetti (2000) dust extinction law (allowing reddening 
to range up to $A_{V}=\rm6$).  This conservative set of parameters allows highly-reddened low-redshift 
solutions to be considered in addition to any potential high-redshift solution (Dunlop et al. 2007).

Photometric redshift determinations were performed with and without the 
$K_{S}$ and IRAC photometry.  Objects that had a high-redshift ($z\geq8.4$) primary solution 
in both runs were retained.  Such objects were then visually inspected in all images to check 
that they were real, and that they were not marginally detected in any optical image or optical image stack.  
Although the inclusion of $K_{S}$-band and IRAC fluxes helped distinguish between low and high-redshift solutions, 
in practice the primary photometric redshift of the high-redshift candidates was generally 
unaffected by the exclusion or inclusion of the long-wavelength data.  
This is as expected, given that the derived value of $z_{phot}$ at high-redshift is primarily driven by the position of the Lyman break.

\section{High-redshift Galaxy candidates}
In this section we list all of the robust $z>8.4$ galaxy candidates that were found in our 
search of the CLASH and HFF datasets.  The basic measured and derived properties of the 
galaxies in the final sample are provided in Tables 1 \& 2.  First we discuss the HFF candidates, 
including those from the two new data releases.  We then discuss the candidates found in the 
twenty CLASH fields analysed in this study.  For a more detailed comparison with previous 
searches in the first two HFF datasets, the reader is referred to McLeod et al. (2015).

\subsection{HFF: Abell 2744 and MACS0416.1-2403 cluster and parallel fields}
In McLeod et al. (2015) we found twelve galaxies in the redshift range $8.4\leq z_{phot}<9.5$ 
in the cluster and parallel fields of Abell2744 and MACS0416.  However, the IRAC 
photometry was not included in the SED fitting performed in McLeod et al. (2015).  
A re-analysis of these fields using both $K_{S}$ and IRAC data
recovers all of these galaxies in the redshift range of interest.  
Notably, one extra candidate is included from the MACS0416 parallel
field.  HFF2P-9-5 was found in the McLeod et al. (2015) study, but was deemed too insecure to be included in the final sample.  
However, thanks to the inclusion of the longer wavelength IRAC photometry, this candidate now has a 
secure primary redshift solution of $z_{phot}$=$9.1^{+0.7}_{-0.5}$.

\subsection{HFF: MACS J0717.5+3745 cluster and parallel fields}
The imaging of the MACS J0717.5+3745 field constituted the third data
release from the HFF programme. A study of the cluster field by Kawamata et
al. (2015) produced no $z\sim 9$ galaxy candidates. Most recently, a
study of the cluster and parallel field by Laporte et al. (2016) also revealed no $z\sim 9$ candidates.
Our new, independent, analysis also reveals no credible $z\geq9$
galaxy candidates in either the cluster or the parallel field.  As noted previously, the effective area of the 
MACS0717 cluster field is much lower than that of the other HFF fields, due to 
severe source crowding which reduces the depth over much of the image.

\subsection{HFF: MACS J1149+2223 cluster and parallel fields}
The MACS J1149+2223 cluster field was previously studied as part of
the CLASH programme and yielded one robust $z\simeq9$ galaxy
candidate, which was identified in the analysis of both Zheng et al. (2012) and Bouwens et al. (2014).  
In their more recent analysis of the much deeper HFF imaging of the
MACS1149 cluster field, Kawamata et al. (2015) identified four
$z\simeq9$ candidates, one of which was the original object identified
from the CLASH dataset.

Our analysis of the MACS1149 cluster field has identified two $z\simeq
9$ galaxy candidates. The first (HFF4C-9-1, $z_{phot}=9.3$) is the
original CLASH candidate and the second (HFF4C-9-2, $z_{phot}=8.7$) 
is a candidate which also identified by Kawamata et
al. (2015). Consequently, there are two $z\simeq 9$ candidates
identified by Kawamata et al. (2015) in the MACS1149 cluster field
which do not feature in our final sample. The first of these
(HFF4C-Y2 in Kawamata et al.) is included in our original
catalogues, but is located in a region of the HFF imaging which is too
shallow to be included in our blank-field search area. Our primary
photometric redshift solution for this object is
$z_{phot}=1.1^{+7.2}_{-0.2}$. However, the very flat $\chi^{2}-z$
distribution means that we cannot rule-out the possibility that this object is at high redshift. 
We cannot rule-out the possibility that the second object (HFF4C-YJ4) is at high redshift either. This object does not feature in our final high signal-to-noise
catalogues, simply because it lies close to the cluster core in a region of shallow depth and large (but uncertain) magnification.

Our analysis of the HFF imaging of the MACS1149 parallel field
identifies two new candidates; HFF4P-9-1 at $z_{phot}=9.0$ and
HFF4P-10-1 at $z_{phot}=9.5$. To date, there are no other analyses
of the HFF imaging of the MACS1149 parallel field in the literature.

\subsection{The CLASH survey}
In their study of the CLASH dataset, Bouwens et al. (2014) found a single $z\simeq9-10$ candidate in each of MACS1115 and MACS1720 clusters.  
Our new analysis of the CLASH dataset has produced a final sample of
15 galaxy candidates in the redshift range $8.4<z_{phot}<11.2$,
including the two candidates previously identified by Bouwens et al. (2014).

Of the thirteen candidates in our final sample which were not
identified by Bouwens et al. (2014), four have been selected from
CLASH clusters which were not analysed by Bouwens et al. (2014).
The remaining nine objects either have colours which just fail to
satisfy the Bouwens et al. selection criteria, or have
photometric redshifts outside the $z \simeq 9.2\pm0.5$ range where
the Bouwens et al. colour-colour selection is designed to operate.
 
We note that our final CLASH sample features one object at
$z_{phot}=11.2$ (M0647-11-1). This object was first discovered by Coe
et al. (2013) as JD1.  In that study, Coe et al. identified two
possible counter images of this object; JD2 and JD3. Based on our photometry, we could not confirm the high-redshift solution for JD2, and so it is not included here.  Our SED
fitting analysis of JD3 produces a primary high-redshift solution at
$z=10.3^{+0.5}_{-0.3}$ and this object is listed in Table 2 as M0647-10-1.

\begin{table*}
 \caption{The final $z\geq 8.4$ galaxy sample in the Hubble Frontier Fields.
Column one lists the candidate names, while columns two and 
three list their coordinates. Column four gives  the best-fitting
photometric redshift with corresponding uncertainty. Column five gives
the total observed $H_{160}$ magnitude, measured using a 0.85-arcsec
diameter circular aperture, plus a point-source correction for flux
outside the aperture. Column six gives the median source magnification
for the galaxies in the cluster fields -- these are estimated given
the CATS, Zitrin LTM and Zitrin NFW magnification models for the first
two cluster fields.  As the latter is unavailable for MACS0717 and
MACS1149, only the CATS and LTM models were used.  The errors are the
upper and lower extremes between these values.  Columns seven and
eight give the total apparent magnitude in $H_{160}$ and the total absolute magnitude at 1500\AA\,, both after demagnification.  Column nine makes reference to other studies which have independently discovered our candidates: (1) Zheng et al. (2012), (2) Zitrin et al. (2014), (3) Bouwens et al. (2014), (4) Zheng et al. (2014), (5) Coe, Bradley \& Zitrin (2015), (6) Ishigaki et al. (2015), (7) McLeod et al. (2015), (8) Kawamata et al. (2015).}

\begin{tabular}{lccccccccc}
\hline
Candidate ID & RA(J2000) & Dec(J2000) & $z_{phot}$ & $H_{160}$         & $\rm \mu$              &  De-mag $H_{160}$ & $M_{1500}$ &References\\
\hline

HFF1C-9-1 & 00:14:24.93 & $-$30:22:56.15 & 8.4$^{+0.3}_{-0.2}$  & 26.75$^{+0.08}_{-0.08}$& 1.5$^{+0.6}_{-0.1}$ & 27.2$^{+0.1}_{-0.1}$& $-$19.9   & 4, 5, 6, 7 \\[1ex]

HFF1C-10-1& 00:14:22.80 & $-$30:24:02.71 & 9.5$^{+0.9}_{-9.0}$ & 27.00$^{+0.24}_{-0.19}$ & 13.2$^{+1.8}_{-1.2}$ & 29.8$^{+0.2}_{-0.2}$ & $-$17.6 & 2, 4 \\[1ex]

HFF1P-9-1 & 00:13:57.33 & $-$30:23:46.27 &  8.8$^{+0.7}_{-0.2}$   & 27.91$^{+0.17}_{-0.15}$ & -                & 27.9$^{+0.2}_{-0.2}$& $-$19.5   & 6, 7           \\[1ex]

HFF1P-9-2 & 00:13:53.64 & $-$30:23:02.48 & 9.3$^{+0.5}_{-0.6}$ & 28.04$^{+0.24}_{-0.20}$& -                & 28.0$^{+0.2}_{-0.2}$& $-$19.3   & 6, 7          \\[1ex]

HFF2C-9-1 & 04:16:09.40 & $-$24:05:35.46 & 8.6$^{+0.1}_{-0.1}$  & 26.01$^{+0.05}_{-0.05}$& 1.7$^{+0.1}_{-0.3}$ & 26.6$^{+0.1}_{-0.1}$& $-$20.6   & 7 \\[1ex]

HFF2C-9-2 & 04:16:11.52 & $-$24:04:54.00 & 8.5$^{+0.1}_{-0.2}$   & 26.55$^{+0.04}_{-0.04}$& 1.7$^{+0.2}_{-0.1}$ & 27.1$^{+0.1}_{-0.1}$& $-$20.2   & 5, 7 \\[1ex]

HFF2C-9-3 & 04:16:10.35 & $-$24:03:28.49 & 8.7$^{+1.1}_{-0.5}$  & 27.82$^{+0.17}_{-0.15}$& 3.3$^{+0.1}_{-0.3}$ & 29.1$^{+0.2}_{-0.2}$& $-$18.1   & 7\\[1ex]

HFF2C-9-4 & 04:16:09.02 & $-$24:05:17.18 & 8.5$^{+0.1}_{-7.5}$ & 27.92$^{+0.27}_{-0.21}$& 1.9$^{+0.5}_{-0.3}$ & 28.6$^{+0.3}_{-0.2}$& $-$18.7   & 7\\[1ex]

HFF2C-9-5 & 04:16:11.09 & $-$24:05:28.82 & 8.6$^{+0.9}_{-0.6}$ & 28.03$^{+0.28}_{-0.23}$& 1.4$^{+0.2}_{-0.1}$ & 28.4$^{+0.3}_{-0.2}$& $-$18.9   & 7\\[1ex]

HFF2P-9-1 & 04:16:35.97 & $-$24:06:48.08 & 8.7$^{+0.7}_{-0.3}$  & 27.69$^{+0.11}_{-0.10}$& -                & 27.7$^{+0.1}_{-0.1}$& $-$19.6   & 5, 7 \\[1ex]

HFF2P-9-2 & 04:16:31.72 & $-$24:06:46.77 &  9.3$^{+0.3}_{-0.6}$ & 28.37$^{+0.20}_{-0.17}$& -                & 28.4$^{+0.2}_{-0.2}$& $-$19.1   & 7\\[1ex]

HFF2P-9-3 & 04:16:36.43 & $-$24:06:30.44 & 8.9$^{+0.5}_{-0.3}$ & 28.11$^{+0.17}_{-0.15}$& -                & 28.1$^{+0.2}_{-0.2}$& $-$19.3   &5, 7 \\[1ex]

HFF2P-9-4 & 04:16:30.43 & $-$24:06:01.13 & 8.4$^{+0.3}_{-0.4}$ & 28.08$^{+0.14}_{-0.12}$& -                & 28.1$^{+0.1}_{-0.1}$& $-$19.1   & 5, 7 \\[1ex]

HFF2P-9-5 & 04:16:36.40 & $-$24:06:29.63 & 9.1$^{+0.7}_{-0.5}$ & 28.32$^{+0.21}_{-0.18}$ & - & 28.3$^{+0.2}_{-0.2}$ &$-$19.0 & \\[1ex]

HFF4C-9-1 & 11:49:33.59 & $+$22:24:45.76 & 9.3$^{+0.3}_{-0.4}$ & 25.58$^{+0.02}_{-0.02}$ & 8.7$^{+0.1}_{-0.1}$ & 27.9$^{+0.1}_{-0.1}$ & $-$19.4 & 1, 3, 8\\[1ex]

HFF4C-9-2 & 11:49:33.73 & $+$22:24:48.30 & 8.7$^{+1.2}_{-8.4}$ & 27.98$^{+0.29}_{-0.23}$ & 9.2$^{+1.4}_{-1.4}$ & 30.4$^{+0.3}_{-0.2}$ & $-$16.9 & 8\\[1ex]

HFF4P-9-1 & 11:49:37.33 & $+$22:17:35.35 & 9.0$^{+0.9}_{-0.5}$ & 28.32$^{+0.35}_{-0.27}$ & - &  28.3$^{+0.4}_{-0.3}$ &$-$19.0 & \\[1ex]

HFF4P-10-1& 11:49:39.95 & $+$22:17:36.74 & 9.5$^{+0.2}_{-0.6}$ & 27.49$^{+0.15}_{-0.13}$ & - & 27.5$^{+0.2}_{-0.1}$ &$-$19.9 & \\[1ex]
\hline
\end{tabular}

\end{table*}
 \begin{table*}
 \caption{The final $z\geq 8.4$ galaxy sample in the twenty CLASH fields.
Column one lists the candidate names, while columns two and 
three list their coordinates. Column four gives the best-fitting photometric redshift along with the corresponding uncertainty. Column five gives the total observed $H_{160}$ magnitude, measured
using a 0.85-arcsec diameter circular aperture, plus a point-source correction for flux outside the aperture. Columns six and seven give the source magnification for the candidates - these are estimated from the Zitrin LTM gauss and Zitrin NFW magnification models.  Column eight gives the absolute magnitude at 1500\AA\, before demagnification.  These can be demagnified with $\mu$ to find the intrinsic $M_{1500}$, which was used in the LF calculations. Column nine makes reference to other studies which have independently discovered our candidates: (1) Coe et al. (2013), (2) Bouwens et al. (2014).  $\ddag$ These candidates are probably the same triply imaged object, as discussed in Coe et al. (2013).}
\begin{tabular}{lccccccccc}

\hline
Candidate ID & RA(J2000) & Dec(J2000) & $z_{phot}$ & $H_{160}$         & $\rm \mu_{ltm-gauss}$              &  $\rm \mu_{nfw}$ & Magnified $M_{1500}$ &References\\
\hline
  A383-9-1 & 02:48:03.25 & $-$03:32:45.60 & \,9.0$^{+0.5}_{-0.3}$ & 27.55$^{+0.71}_{-0.43}$ & \,1.4 & \,1.4 & $-$19.8\\[1ex]
  
  A209-9-1 & 01:31:54.30 & $-$13:36:57.11 & \,8.4$^{+1.0}_{-7.6}$ & 26.17$^{+0.29}_{-0.23}$ & \,2.5 & \,4.8 & $-$21.1\\[1ex]
  
  A611-10-1 & 08:01:01.08 & $+$36:03:03.67 & \,9.9$^{+0.4}_{-0.5}$ & 27.12$^{+0.40}_{-0.29}$ & \,3.0 & \,1.5 & $-$20.3\\[1ex]
  
  M0429-10-1 & 04:29:37.79 & $-$02:54:02.26 & \,9.6$^{+0.6}_{-0.9}$ & 27.33$^{+0.47}_{-0.33}$  & \,3.1 & \,1.7 & $-$20.1\\[1ex]
  
  M0647-11-1$^{\ddag}$ & 06:47:55.73 & $+$70:14:35.76 & 11.2$^{+0.3}_{-0.5}$ & 25.46$^{+0.11}_{-0.10}$ & \,5.6 & \,5.6 & $-$22.2 & 1\\[1ex]
  
  M0647-10-1$^{\ddag}$ & 06:47:55.46 & $+$70:15:38.08 & 10.3$^{+0.5}_{-0.3}$ & 26.66$^{+0.17}_{-0.15}$ & \,3.2 & \,1.9 & $-$20.9 & 1\\[1ex]
  
  M0647-9-1 & 06:47:47.76 & $+$70:15:41.98 & \,8.8$^{+0.7}_{-0.8}$ & 27.43$^{+0.42}_{-0.30}$ & \,3.1 & \,2.1 & $-$19.9\\[1ex]
  
  M1115-10-1 & 11:15:54.49 & $+$01:29:47.92 & \,9.6$^{+0.3}_{-0.4}$ & 26.22$^{+0.13}_{-0.12}$ & \,5.7 & \,3.1 & $-$21.2 & 2\\[1ex]
  
  M1115-9-1 & 11:15:54.85 & $+$01:29:27.58 & \,9.1$^{+0.5}_{-0.4}$ & 26.79$^{+0.27}_{-0.22}$ & \,5.6 & \,2.7 & $-$20.5\\[1ex]
  
  M1311-9-1 & 13:10:59.95 & $-$03:10:36.34 & \,8.9$^{+0.7}_{-7.9}$ & 27.50$^{+0.60}_{-0.38}$ & \,6.1 & \,2.9 & $-$19.8\\[1ex]
  
  M1423-9-1 & 14:23:49.08 & $+$24:05:13.68 & \,8.7$^{+0.3}_{-7.8}$ & 27.09$^{+0.40}_{-0.29}$& 17.9 & 20.7 & $-$20.2\\[1ex]
  
  M1423-9-2 & 14:23:45.89 & $+$24:04:09.54 & \,8.5$^{+0.4}_{-0.5}$ & 26.99$^{+0.35}_{-0.27}$ & \,9.7 & \,5.5 & $-$20.2\\[1ex]
  
  M1720-9-1 & 17:20:12.74 & $+$35:36:17.25 & \,9.2$^{+0.4}_{-0.7}$ & 26.73$^{+0.24}_{-0.20}$ & \,2.9 & \,1.3 & $-$20.6 & 2\\[1ex]
  
  M1931-9-1 & 19:31:48.83 & $-$26:33:29.33 & \,8.7$^{+0.6}_{-7.8}$ & 27.26$^{+0.47}_{-0.33}$ & \,2.6 & \,3.5 & $-$20.0\\[1ex]
  
  RXJ2129-9-1 & 21:29:39.91 & $+$00:05:58.79 & \,8.8$^{+0.6}_{-0.5}$ & 26.35$^{+0.16}_{-0.14}$ & \,2.1 & \,1.4 & $-$20.9\\[1ex]
\hline\end{tabular}
\end{table*}

\begin{figure}
\includegraphics[width=6.0cm, angle=270]{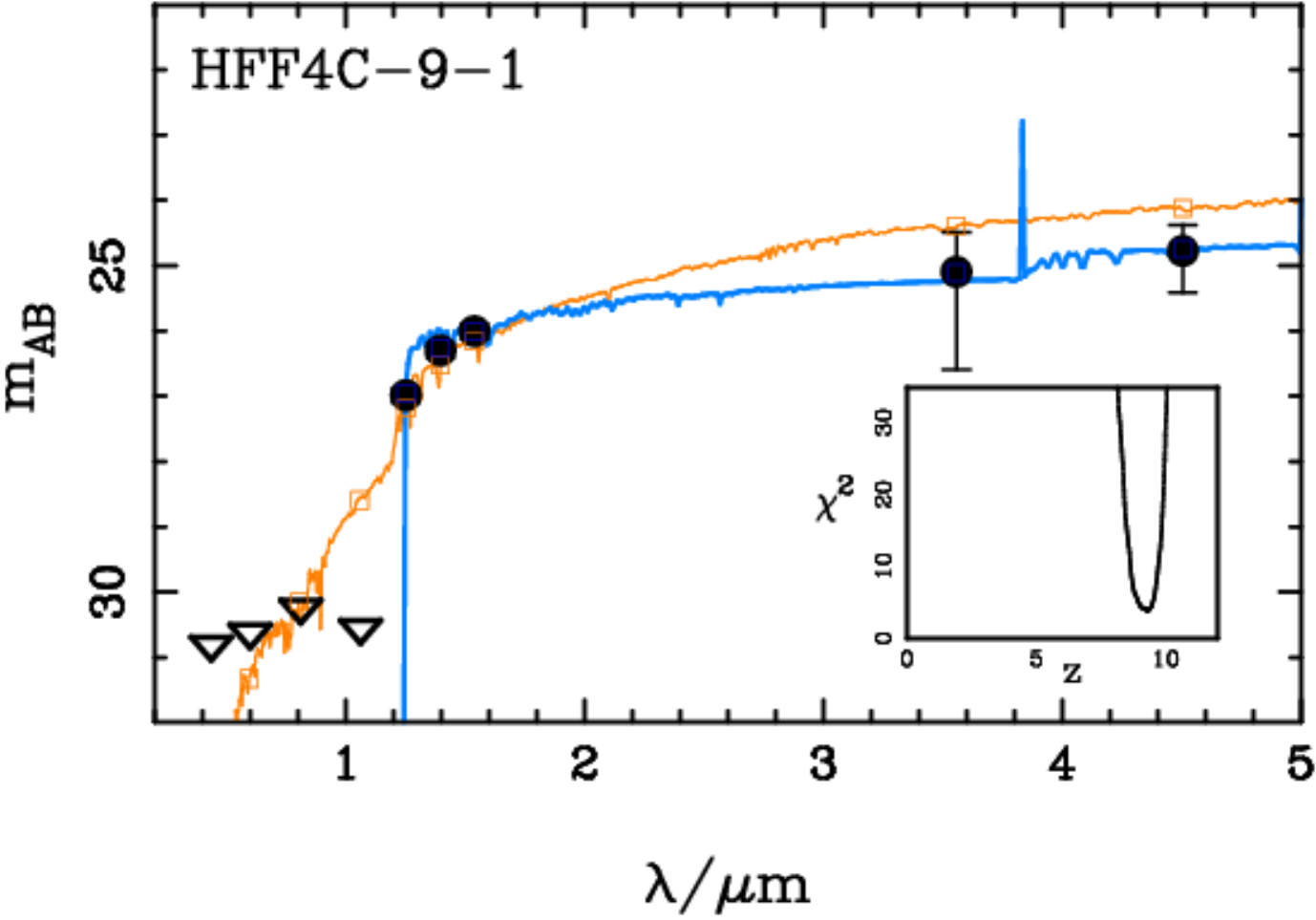}
\caption{The results of our SED fitting for HFF4C-9-1. The candidate
  photometry is shown in black, with non-detections plotted at
  $1\sigma$ upper limits. The best-fitting SED model with $z_{phot}=9.3$ is shown in blue and 
the best alternative solution with $z_{phot}=2.2$ is shown in
orange. The distribution of $\chi^2$ versus redshift is shown in the
inset panel. For this candidate the high-redshift solution is strongly
preferred, with the alternative solutions shown in the figure
separated by $\Delta\chi^2=41$. The small squares show the synthetic
photometry produced by the two alternative SED models.}
\end{figure}

\begin{figure*}
\includegraphics[width=18.5cm, angle=0]{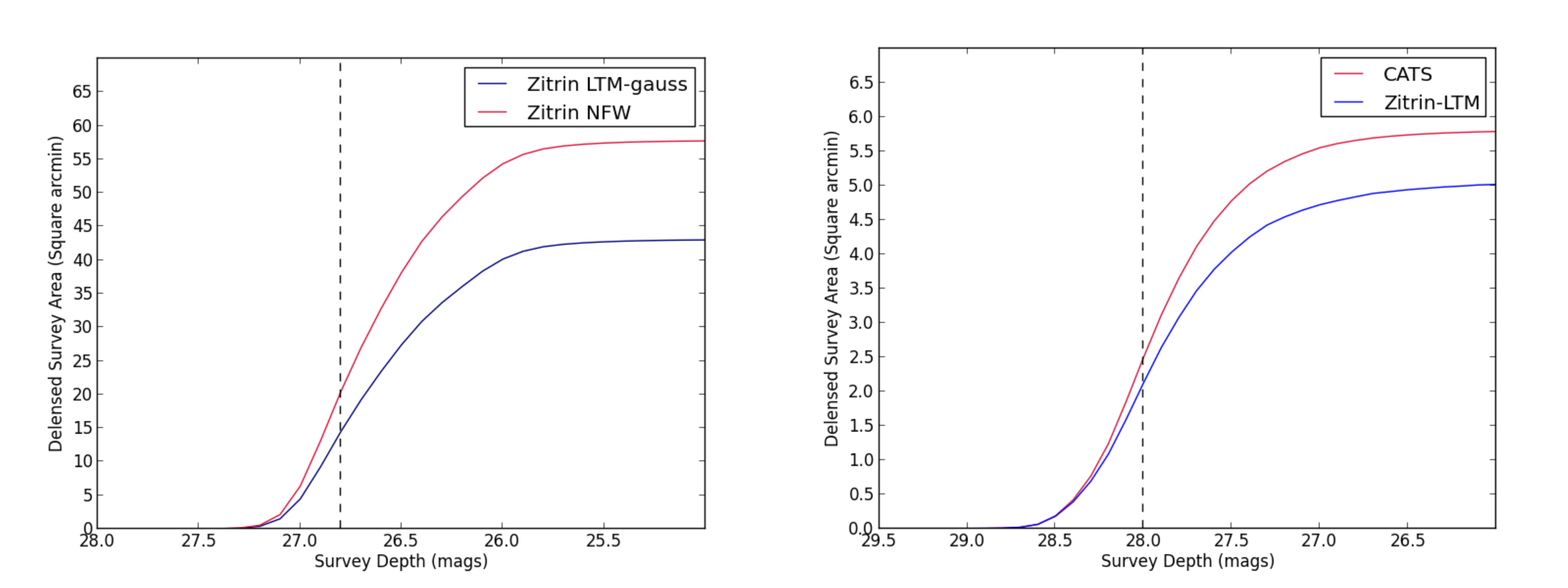}
\caption{Left: The combined, effective, delensed survey area of the twenty CLASH clusters used in this study, 
as a function of the 5$\sigma$ $H_{160}$ depth.  This illustrates the difference between the 
two different Zitrin et al. (2013) lensing models employed in this study.  
Right:  The same, but for the four HFF cluster fields.  
The comparison here is between the two models by CATS and Zitrin et al. (LTM).  For each plot, 
the adopted 5$\sigma$ $H_{160}$ depth limit is indicated by the dashed line; at $H_{160}=28.0$ for the HFF imaging 
and $H_{160}=26.8$ for the CLASH imaging.  Although not shown here, the depth cut-off at $H_{160}=28.0$ serves to 
retain $\simeq$85\% of the total effective area available in the HFF parallel fields.}
\end{figure*}

\section{The Luminosity Function at z=9: Blank Field Method}
\subsection{Effective search area}
\indent By analysing the local depth maps, object selection and effective survey area can be usefully confined 
to only the cleanest and deepest regions of the {\it HST} imaging.  As
in McLeod et al. (2015), we refer to this as the ``blank field''
method for determining the $z=9$ LF. For the HFF the depth cut was chosen to be 28.0 mag 
in $H_{160}$ (at 85\% of total assuming point-source), as in McLeod et al. (2015), again meaning 
that we focus on regions where $\sigma_{160}\simeq30$ mag (or deeper) in a 0.5$^{\prime\prime}$ diameter aperture.  
For the CLASH fields the cut-off was $H_{160}=26.8$ at 85\% of total.

In addition to imaging depth, one also has to consider the reduction in effective area produced by factoring in the magnification maps.  
For the HFF fields analysed here, we have employed the CATS 
(Richard et al. 2014) and Zitrin LTM (Zitrin et al. 2013) lensing maps.  
For the CLASH fields, two lensing maps have been publicly released, the NFW and LTM-gauss (LTMG) models by Zitrin et al. (2013), 
and so the analysis was done using those. In Fig. 4, we plot the effective de-lensed area of our combined CLASH 
imaging as a function of the survey depth for the lensing models adopted.

The lensing maps were also used to calculate the magnification of each candidate.  
For each map, the magnification was based on the average model magnification 
within a 0.5$^{\prime\prime}$ diameter aperture around each object.  
For this blank-field method, the magnification factor was taken as the average of the 
different model values, with the extremes as the error bounds.

Using the aforementioned magnitude cuts, the delensed areas found were 20.4\,arcmin$^{2}$ (NFW) 
and 14.5\,arcmin$^{2}$ (LTMG) for CLASH, with 18.6\,arcmin$^{2}$ for HFF.  For the latter, the area is completely dominated by 
the parallel fields, which together contribute around 16.5\,arcmin$^{2}$ of effective survey area.  
The area adopted for the HFF clusters was the average area from the two lensing maps used.

Completeness was calculated following the prescription described in McLure et al. (2013).  
Simulations in which sources were injected into the imaging and recovered were performed 
in order to find the completeness as a function of magnitude.  
The interested reader is referred to McLeod et al. (2015) for more details.

\subsection{The $\bmath{z=9}$ luminosity function}
Using the $z \simeq 9$ blank-field sample, we are now in a position to better constrain the UV LF at this epoch.  
We find that none of the CLASH objects contribute to any LF determination in the blank field approach.  
This is is because they either lie in regions of the imaging that are shallower than the adopted cut-off limit or, 
when de-magnified, they are intrinsically too faint to be detected if the field is assumed to be a blank one.
Incompleteness prevents us from including the fainter,
$M_{1500}>-18.7$, galaxies uncovered in the HFF pointings, but 
we can extend our study to fainter magnitudes by combining our new HFF results with the HUDF12 results at $z \simeq 9$ 
derived by McLure et al. (2013). This results in  four luminosity bins at $z \simeq 9$, spanning the absolute 
magnitude range $-20.7<M_{1500}<-17.25$. Finally, we did not find any galaxies brighter than $M_{1500}=-20.7$ in our search 
of CLASH and the HFF datasets, and therefore derive an upper limit at $log_{10}\phi=-4.90$ at this magnitude (corresponding 
to a single object in the CLASH+HFF+HUDF volume searched).

These new results are presented in Fig.\,5, which shows the 
$z \simeq 9$ UV LF as determined by restricting our analysis to the
clean regions of the HFF and HUDF imaging, covering an area of $\simeq$ 22 arcmin$^{2}$.  The fainter two bins
come from an area of $\simeq$ 4 arcmin$^{2}$ provided by the HUDF
alone.  The upper limit at the bright end is derived from the unsuccessful search for $z \simeq 9$ 
candidates brighter than $M_{1500}=-20.7$ within a total area of
$\simeq 45$\,arcmin$^{2}$ from the CLASH, HFF and HUDF 
imaging which reached 5$\sigma$ depths greater than 26.8 mag.

As also shown in Fig.\,5, we have fitted a Schechter function (Schechter 1976) to the data.  
Because the data at $z \simeq 9$ still do not provide enough accuracy
and/or dynamic range to allow a robust measurement of the faint-end slope, during the fitting this was held fixed at the $z \simeq 8$ value of $\alpha=-2.02$, 
as determined by McLure et al. (2013). We also cannot yet distinguish whether density or luminosity evolution 
better describes the evolution of the LF from $z \simeq 8$ to $z
\simeq 9$, finding that the $z \simeq 9$ LF can
be well described either by a drop in density by a factor $\simeq 1.8$ to $\log_{10} \phi^{\star}=-3.62^{+0.08}_{-0.10}$
(for a fixed $M^{*}=-20.1$), or by a dimming of the characteristic
magnitude by $\simeq 0.45$ \,magnitudes to $M^{\star}=-19.65\pm0.15$ 
(for a fixed $\log_{10} \phi^{\star}=-3.35$). These two alternative Schechter function fits at $z \simeq 9$ are both shown in Fig.\,5, 
and are essentially indistinguishable from the $z \simeq 9$ LF fits derived and presented by McLeod et al. (2015).

As previously mentioned, these results are derived by adopting the average magnifications of the two alternative
lensing models available for the new HFF fields. However, it can be
seen from Fig.\, 4 that there is very little difference in the derived effective survey 
area between the different HFF lensing models for objects at our
adopted blank-field magnitude limit.

\begin{figure}
\centering
\includegraphics[width=\columnwidth, angle=0]{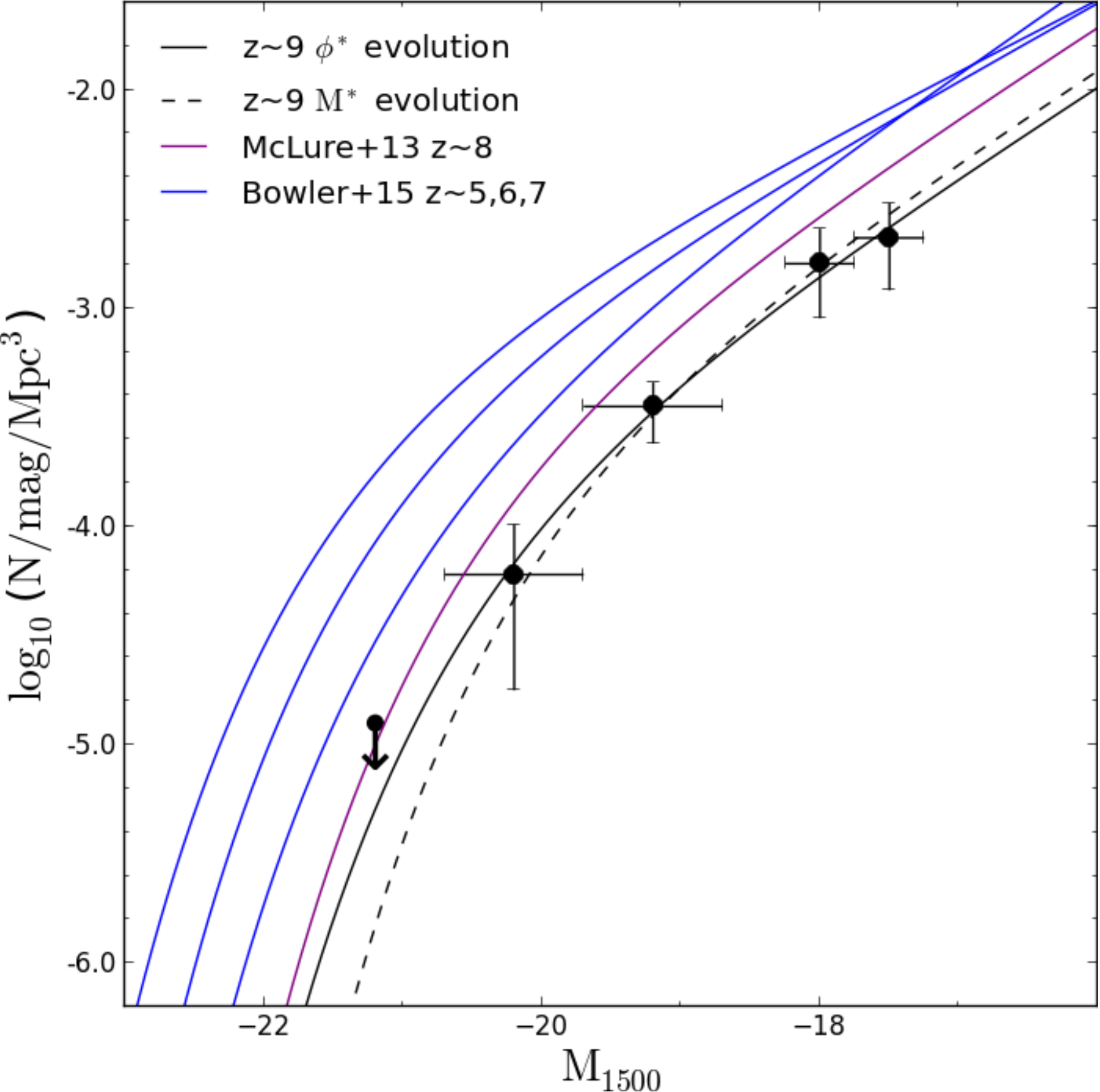}
\caption{Our new determination of the UV LF at $z\simeq9$ via the blank field approach (i.e. constraining our search to the cleanest 
regions of the available HFF and CLASH imaging).  The blue curves are the Schechter function fits to the $z=5$, $z=6$ and $z=7$ LF 
from Bowler et al. (2015), while the purple curve is the $z=8$ LF fit from McLure et al. (2013).  The fainter bins corresponding to 
$M_{1500}=-17.5$ and $M_{1500}=-18.0$ are from the HUDF study of McLure et al. (2013), while the brighter two bins at $M_{1500}=-19.2$ and 
$M_{1500}=-20.2$ are the result of the present HFF study.  An upper limit is included at $M_{1500}=-21.2$ resulting from 
our unsuccessful search for $z \simeq 9$ galaxies at these brighter magnitudes over the $\simeq 45$\,arcmin$^{2}$ 
of combined HUDF+HFF+CLASH imaging of appropriate depth.  Alternative Schechter function fits are shown at $z \simeq 9$, 
one assuming pure luminosity evolution from $z \simeq 8$ (dashed black line), and the other assuming pure number density evolution 
from $z \simeq 8$ (solid black line). The current data do not allow us to distinguish between 
these two simple alternative scenarios. In both of these fits the faint-end slope was set to the value derived by McLure et al. (2013) 
at $z \simeq 8$ (i.e. $\alpha =-2.02$).}
\end{figure}

\section{The Luminosity Function at $\bmath{z=9}$: Lensed Field Method}

\subsection{Lensed areas}

By adopting the blank field method, none of our high-redshift candidates from the CLASH fields feature in 
the $z=9$ LF determination.  To be as inclusive of our CLASH candidates as possible, we attempted to derive a 
luminosity function using the full available area of the CLASH survey, including the more highly-lensed regions.  
This comes with the risk of introducing biases in any derived number densities, due to uncertainties in the 
magnifications.  As the majority of the CLASH candidates have come from a stacked image, 
the analysis was carried out on a $J_{140}+H_{160}$ stacked image for all fields.  For each object, the 
delensed $J_{140}+H_{160}$ flux is compared to the delensed depth map of every cluster field.  
Any pixels that are of sufficient depth that the object would be detectable at 5$\sigma$ in the stack 
contributed to the total available cosmological volume.  Subject to a completeness correction, we are then 
left with the effective volume contribution of each candidate in the sample.  The magnification models were 
treated entirely separately when calculating magnification values and effective volumes.  The delensed area-magnitude plots 
shown in Fig.\,6 illustrate how this modifies the effective area calculations.

\begin{figure}
\includegraphics[width=\columnwidth, angle=0]{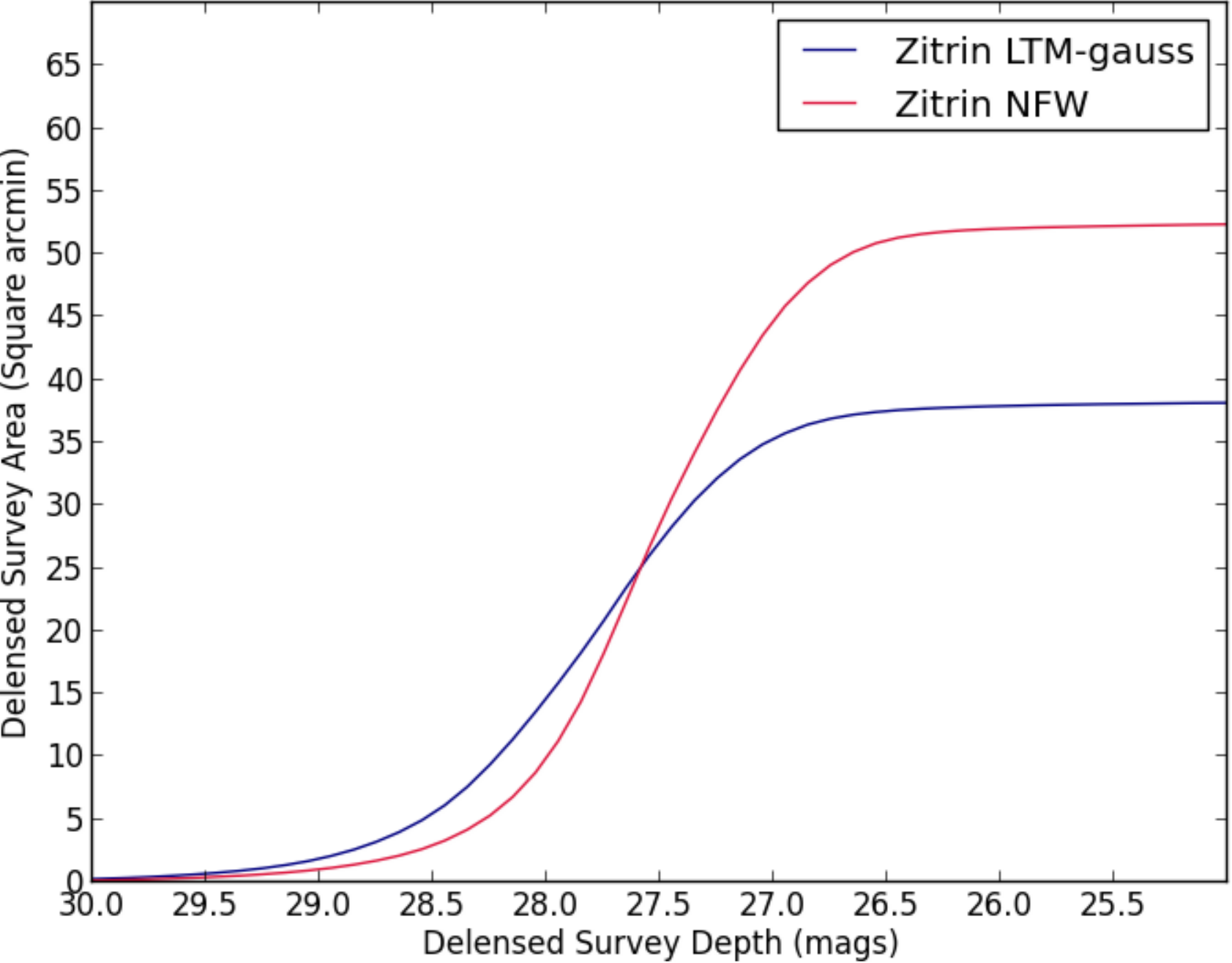}
\caption{The effective area of the twenty CLASH cluster fields as a function of the delensed $J_{140}+H_{160}$ stack depth, for the two magnification models considered.}
\end{figure}

\subsection{Lensed LF calculation}
In Fig.\,7 we present our alternative estimate of the $z=9$ LF based on the lensed CLASH galaxies.  
We plot two alternative LFs, corresponding to the two alternative lensing models.  
Comparing with the blank field LF determination derived earlier, it can be seen that the two methods are generally consistent. 
Although for the lensed case both models yield higher number densities
in the fainter ($M_{1500}=-19.2$) bin, the results can all be comfortably reconciled 
through the large Poissonian uncertainties associated with such small number statistics.
Due to the uncertainties between the models, we treat this alternative LF calculation simply as a consistency check.  
As the blank field LF determination is less systematically biased, we
adopt this as the best estimate of the $z=9$ LF.

\begin{figure}
\centering
\includegraphics[width=\columnwidth, angle=0]{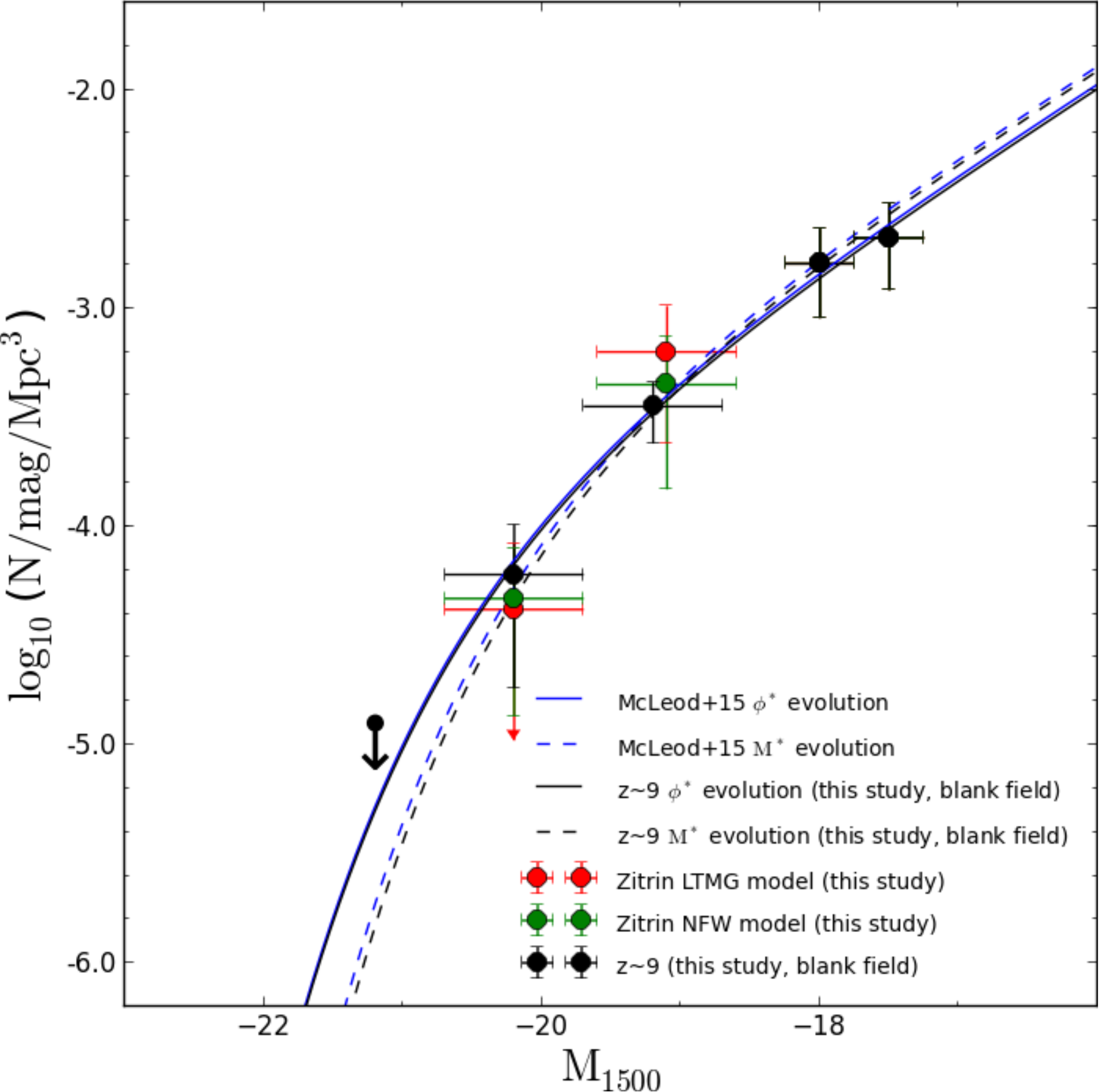}
\caption{Our alternative determination of the UV LF at $z\simeq9$ making fuller use of the search area 
in the CLASH fields.  The black datapoints and curves show the blank-field LF determination from this study as already plotted  
in Fig.\,5, while the coloured datapoints show the binned values resulting from the lensed field studies.
For completeness, the blue curves show the McLeod et al. (2015) $z=9$ fits, illustrating that our new results yield 
essentially identical Schechter function fits at $z = 9$.  
The lensed datapoints correspond to number densities assuming two different magnification models, the Zitrin et al. NFW model (green)  
and the Zitrin LTMG model (red).  The $M_{1500}=-19.2$ lensed datapoints have been offset by +0.1 mag for clarity.  
This plot demonstrates that the results derived using either the blank field or lensed field methodology are very similar, 
but the blank-field approach yields more accurate and robust results.}
\end{figure}

\subsection{Extension to $\bmath{z=10}$}

As the vast majority of our $z\simeq10$ candidates come from the CLASH fields (see Tables 1 and 2), the blank-field approach 
cannot yet be used to determine a meaningful estimate of galaxy number densities at $z\simeq10$.  
Adopting the full-area lensed approach, depending on the assumed lensing model, we have either five candidates (NFW model) or 
three candidates (LTMG model) that can be placed in an absolute magnitude bin centred at $M_{1500}=-19.7$ at $z \simeq 10$.  
One of the candidates comes from the HFF (HFF4P-10-1), with the rest coming from the CLASH survey. The overall 
search area contribution from the HFF, and the number density contribution from the sole HFF candidate, 
were found to be very similar whatever magnification model was adopted, in part due to the dominance of the parallel fields.  
In the following we have adopted the CATS magnification model for the HFF data, but the result is very similar if we substitute this for the Zitrin et al. LTM model.

We present our resulting alternative estimates of galaxy comoving number density at $z\simeq10$ in a single luminosity bin in Fig.\,8. 
This figure also shows our $z \simeq 9$ results, as well as other recent comoving number densities derived at $z \simeq 10$ by 
Oesch et al. (2014). We have fitted a Schechter function through the average of our $z \simeq 10$ datapoints, again fixing $\alpha$ at $-2.02$ 
and assuming either pure $\phi^{\star}$ or pure $M^{\star}$ evolution (fixing the other parameter as appropriate).  
For density evolution, we find that $\phi^{\star}$ evolves in a similar manner between $z=9$ and $z=10$ as between $z=8$ and $z=9$, 
with a further inferred decrease of a factor $\simeq 2$ to $\log_{10}\phi^{\star}$=$-$3.90$^{+0.13}_{-0.20}$. For luminosity evolution, 
we find that the $M^{\star}$ dims by $\simeq$ 0.25 mag from $z \simeq 9$ to $z \simeq 10$,  to $M^{\star}=-19.41 \pm 0.17$.  
Although it is clearly difficult to draw detailed conclusions from a single data point, 
the LF at $z\simeq10$ appears to reflect a smooth extrapolation of the evolution found between $z = 8$ and $z = 9$.

\begin{figure}
\centering
\includegraphics[width=\columnwidth, angle=0]{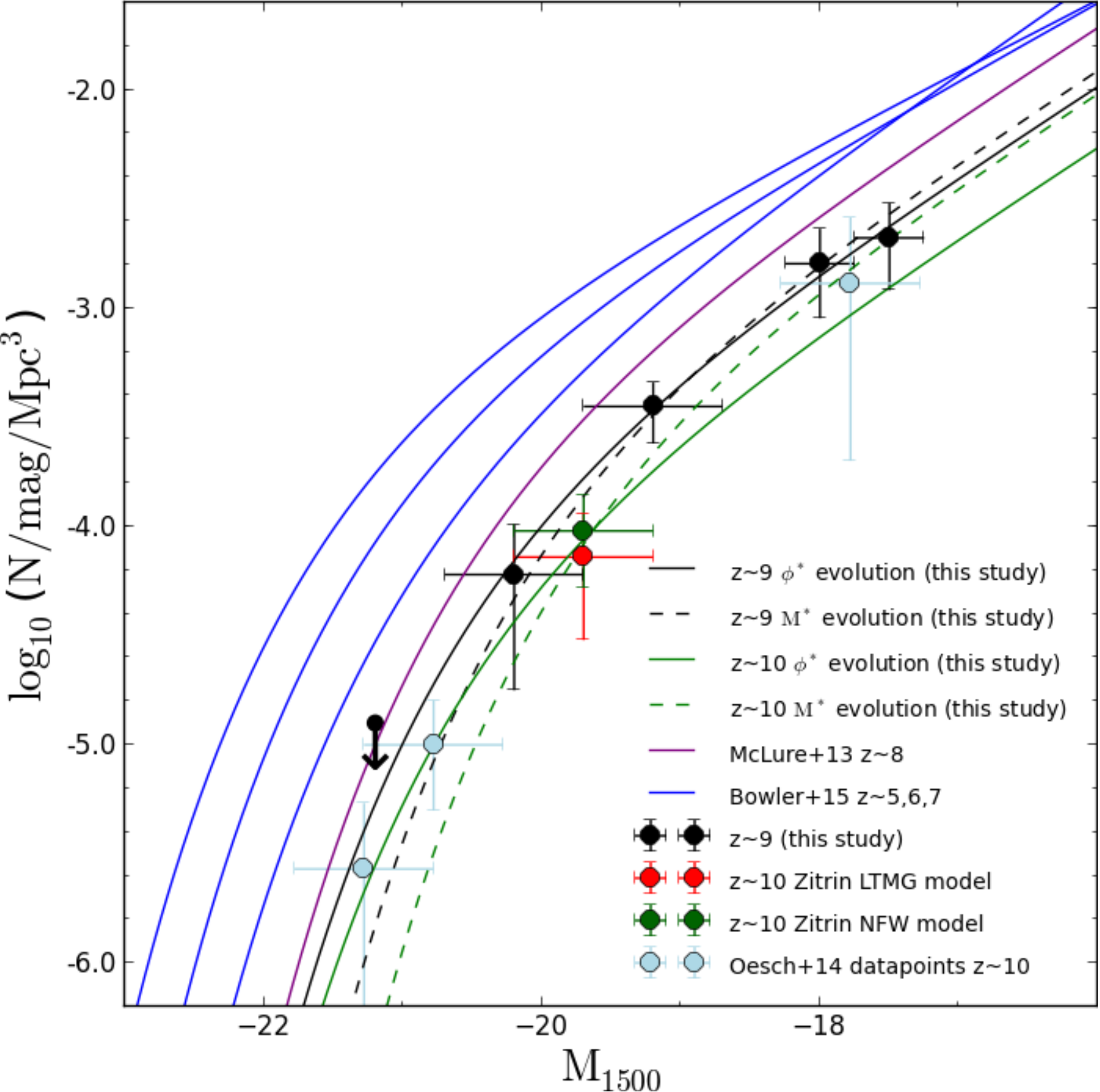}
\caption{This plot shows the same information as shown in Fig.\,5, but now includes our estimate 
of $\phi$ at $M_{1500}=-19.7$ at $z\simeq10$.  The Schechter function fit is again fitted in two ways 
while locking the faint-end slope at the $z \simeq 8$ value - one assuming pure luminosity evolution from $z \simeq 9$, and 
the other assuming pure number density evolution from $z \simeq 9$.  The two alternative 
$z=10$ data points result from adopting either the Zitrin et al. NFW lensing model or the Zitrin et al. LTMG lensing model.  
The inferred Schechter function was found to evolve very similarly between $z=9$ and $z=10$ as between $z=8$ and $z=9$, 
with either a further decline in characteristic number density by a factor of two, 
or a dimming in $M^{\star}$ by a further  $\simeq 0.25$ mag.  The $z=10$ results from Oesch et al. (2014) are also plotted.  
These points appear consistent with our new estimate of number density at $M_{1500}=-19.7$.  
Although the Schechter function was not fitted to these points the agreement with our 
pure number density evolution parameterisation at the bright end is good, although the faintest 
Oesch et al. datapoint marginally favours luminosity evolution and/or further steepening of the faint-end slope.}
\end{figure}

\begin{figure}
\centering
\includegraphics[width=\columnwidth, angle=0]{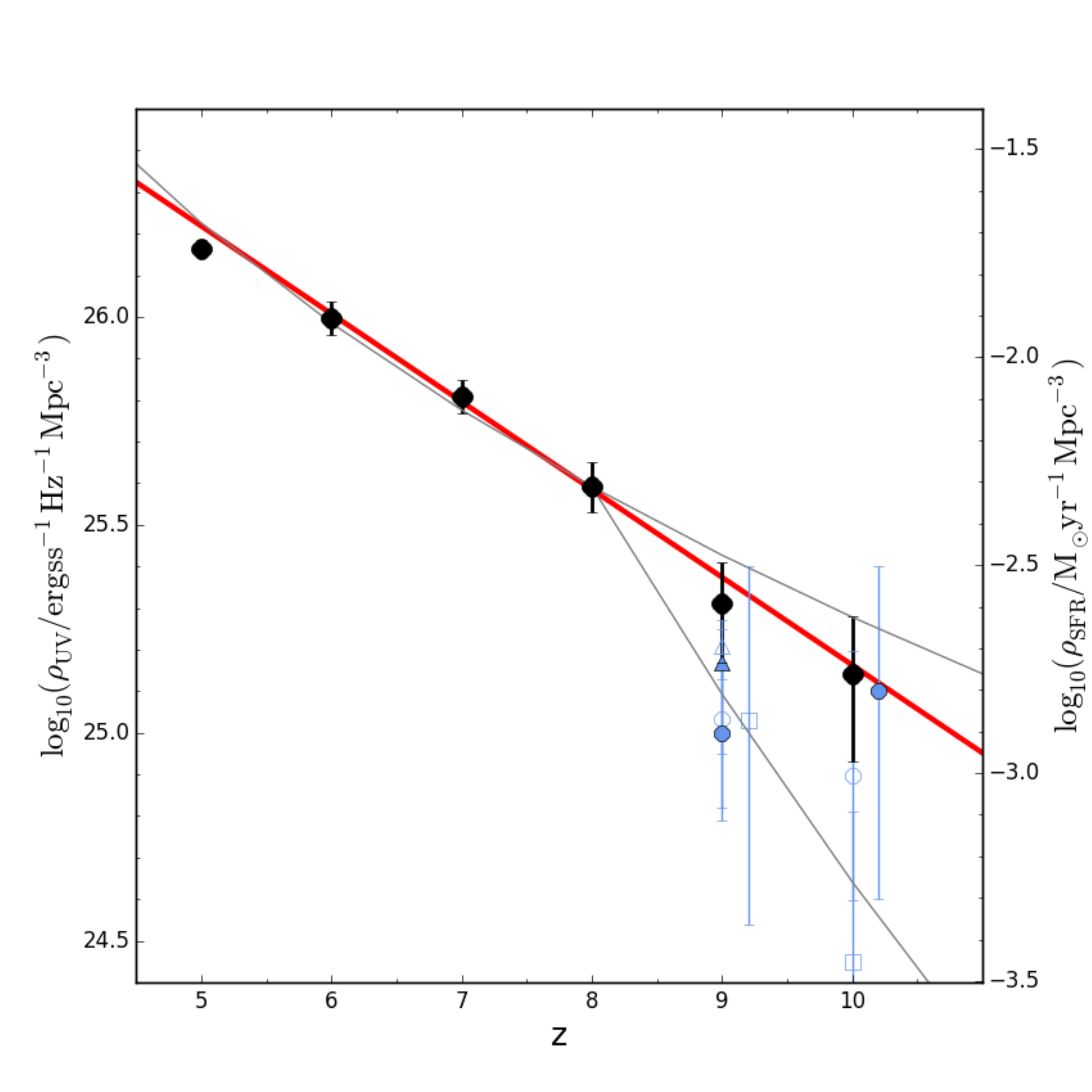}
\caption{The evolution of the UV luminosity density, $\rho_{UV}$, and hence inferred star-formation rate density, $\rho_{SFR}$, at $ z > 5$,
including our new measurements at $z=9$ and $z=10$.  At $z=9$ and $z=10$, the results of several other recent studies are also plotted - open circles, 
filled circles, open triangles, filled triangles and open squares correspond to results from 
Ellis et al. (2013), Oesch et al. (2014, 2015), McLure et al. (2013), Ishigaki et al. (2015) and Bouwens et al. (2015) respectively.  
Our measurements come from integrating the luminosity-weighted Schechter function LFs presented in Fig.\,8 down to $M_{1500}=-17.7$, 
to facilitate straightforward comparison with the other results from the literature.  
The secondary y-axis gives the star-formation rate density, $\rho_{SFR}$, which is derived from $\rho_{UV}$ assuming the 
conversion of Madau, Pozetti and Dickinson (1998) with a Salpeter IMF.  Given that our new estimate of $\rho_{UV}$ at $z \simeq 9$ is
so similar to that of McLeod et al. (2015), we have not re-plotted our earlier result.  
Overplotted in grey is the power-law function $\rho_{UV}\propto(1+z)^{-3.6}$ which is one possible 
simple extrapolation of $\rho_{UV}$ beyond $z>6$, and the steeper evolution following $\rho_{UV}\propto(1+z)^{-10.9}$, as both 
proposed by Oesch et al. (2014).  As our data points do not conform to either of these extreme alternatives, 
we investigated what power-law index would give the best fit, or whether a linear fit to the data would fare better.  
As shown by the red line, 
we find that a linear relation of the form $\log_{10}(\rho_{UV})= -0.211(\pm0.028)z + 27.273(\pm0.193)$ 
provides a good fit to our new estimates of $\rho_{UV}$, and is also in agreement with the $z=10$ luminosity 
density of Oesch et al. (2015).  An alternative is a power-law of the form
$\rho_{UV}\propto(1+z)^{-5.8}$ which is also consistent with our new
estimates (not shown).}
\end{figure}

\section{Evolution of luminosity density at $\bmath{z>8}$}

Given our improved measurement of the galaxy UV LF at $z = 9$, and our
new estimate of the likely evolution to $z \simeq 10$,  we now revisit
the evolution of UV luminosity density ($\rho_{UV}$) at these early epochs. As discussed in the introduction, this is currently the subject of some debate and controversy, 
given the conclusion reached by several recent studies (e.g. Oesch et al. 2014; Bouwens et al. 2015), that there is a dramatic fall-off in $\rho_{UV}$ at $z\geq8$.  

To facilitate comparison with these recent results, we perform the luminosity-weighted integration of 
the best-fitting Schechter functions shown in Fig.\,8 (adopting the luminosity-evolved fits indicated by the dashed lines) down to a faint-end magnitude limit of
$M_{1500}=-17.7$. When performing this calculation we do not dust
correct the data, meaning that we are deliberately calculating the
{\it observed} UV luminosity density. However, it should be noted that
at $z\geq 7$, dust correcting using standard techniques (i.e. based on
the average UV slopes) has a negligible impact on the derived values
of $\rho_{UV}$. As can be seen in Fig.\,9, we find $\log_{10}(\rho_{UV}/ergs\,s^{-1}
Hz^{-1} Mpc^{-3}) = 25.31^{+0.10}_{-0.14}$ at $z \simeq 9$, and $\log_{10}\rho_{UV}=25.14^{+0.14}_{-0.21}$ at $z \simeq 10$.  
We convert this into a comoving star-formation rate density,
$\rho_{SFR}$, by assuming the conversion of Madau, Pozetti and
Dickinson (1998) and a Salpeter IMF, which yields
$\log_{10}\rho_{SFR}=-2.59^{+0.10}_{-0.14}$ at $z \simeq 9$ and
$\log_{10}\rho_{SFR}=-2.76^{+0.14}_{-0.21}$ at $z \simeq 10$ (as also
shown in Fig.\,9, reading from the right-hand axis).

It is clear from Fig.\,9 that our new determinations lie between the two extreme alternative extrapolations of the evolution of  $\rho_{SFR}$ beyond $z \simeq 8$ 
discussed by Oesch et al. (2014). Instead, our new results provide further support for the smooth 
decline of $\rho_{UV}$ to high redshifts as deduced by McLeod et
al. (2015), which can be reasonably described via a linear relation
between $\log_{10}\rho_{UV}$ and $z$. With our new expanded dataset, 
the best fitting linear relation at $z\geq6$ is given by
$\log_{10}(\rho_{UV})= -0.211(\pm0.028)z + 27.273(\pm0.193)$, and this
is plotted as the red line in Fig.\,9. Alternatively, if one wants to
adopt a power-law relation in $(1+z)$, then we find that
$\propto(1+z)^{-5.8}$ also provides a 
good description of the decline in $\rho_{UV}$ beyond $z \simeq 8$.

\begin{figure}
\centering
\includegraphics[width=\columnwidth, angle=0]{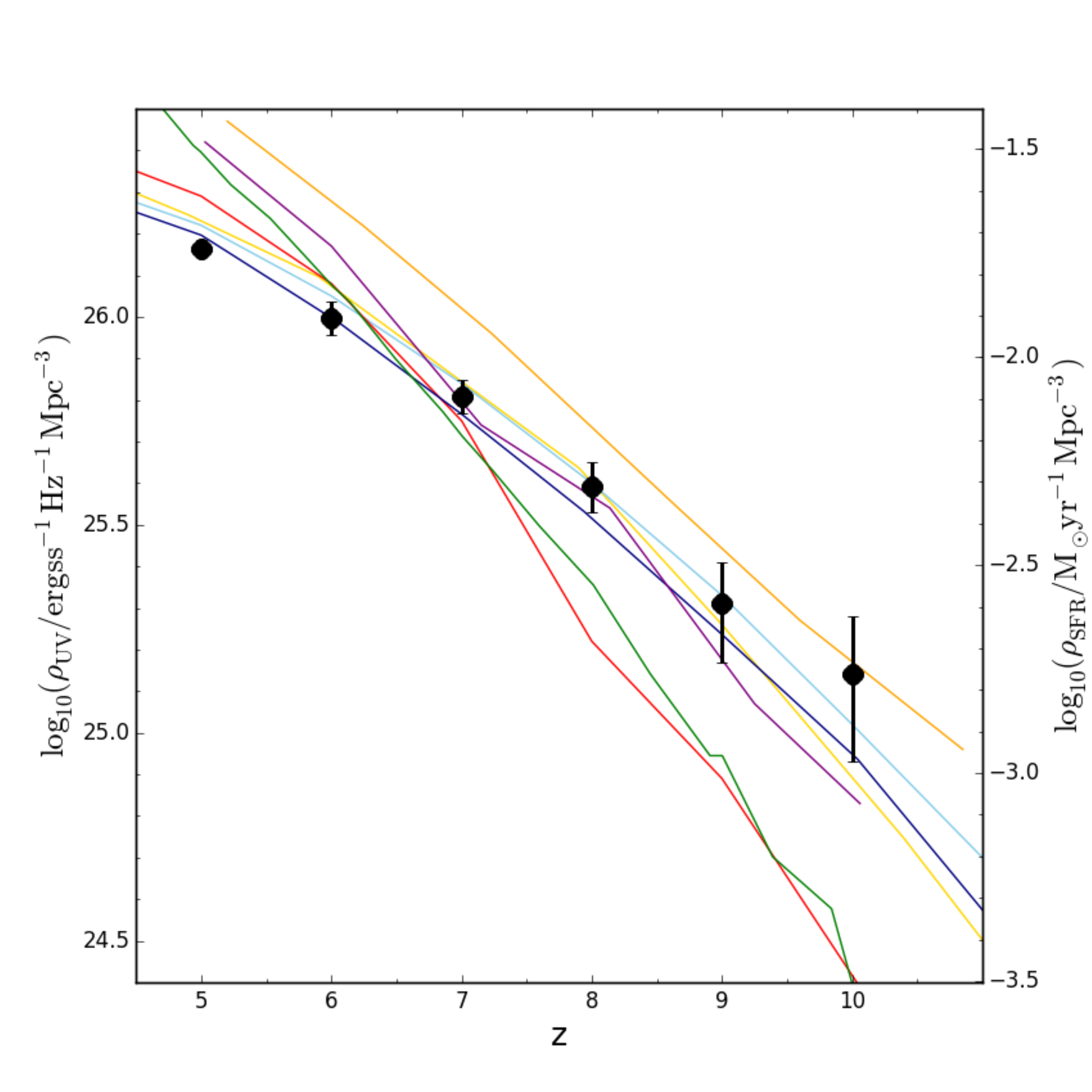}
\caption{Our new measurement of the evolution of the UV luminosity
  density ($\rho_{UV}$) and inferred star-formation rate density ($\rho_{SFR}$) at $z>5$, compared with the predictions of a range of models and
simulations from the literature. The model predictions are as follows:
the Illustris hydro-sim (Genel et al. 2014, green curve), Khochfar et al.,
in preparation, hydro-sim (purple curve), Behroozi \& Silk (2015)
analytic model (orange curve) and the semi-analytic models from Munich
(Henriques et al. 2015) (red curve), Cai et al. (2014) (light blue curve), 
Dayal et al. (2015) (dark blue curve) and Mason et al. (2015) (yellow
curve). Both the Illustris and Munich models under-predict 
the high-redshift luminosity density; this discrepancy is already clear by $z \simeq 8$ and is simply confirmed by our new results at $ z \simeq 9$ and $z \simeq 10$.
By contrast, most of the simple models which basically apply scaling relations to map the underlying halo mass function onto the 
UV LF deliver predictions in reasonably good agreement with our results.}
\end{figure}

Finally, it is interesting to compare our new observational estimate of the high-redshift evolution 
of  $\rho_{UV}$ with the predictions of recently-published models of
galaxy evolution. 

Our data-points
are plotted against the predictions of seven such models in Fig.\,10.
It is important to acknowledge that some of these predictions are at least partly based on hydrodynamic simulations, while others are semi-analytic models, some of which have been effectively 
tuned to reproduce the UV LF at moderately high redshift (e.g. $z \simeq 6$). Thus, in a sense, this 
comparison is not completely fair, and cannot be simply regarded as specifically favouring one form
of model over another. Nevertheless, some important general trends can be deduced. In particular, the 
Illustris (Genel et al. 2014) and Munich (Henriques et al. 2015)
models, while clearly capable of reproducing many (often quite
complex) aspects of galaxy evolution, both clearly under-predict  $\rho_{UV}$ at 
redshifts higher than $z \simeq 7$. We stress that this discrepancy is already clear by $z \simeq 8$ (where the measured value of $\rho_{UV}$ is well established) and is simply confirmed 
and strengthened by our new results at $ z \simeq 9$ and $z \simeq
10$. The challenge, then, for such models is to be able to produce
star-formation activity which extends to higher redshifts than
currently predicted. Indeed, the only model shown here that
systematically over-predicts $\rho_{UV}$ is that of Behroozi \& Silk
(2015). We note that the over-prediction of raw observed 
$\rho_{UV}$ at redshifts $ z \simeq 5 - 6$ by the Illustris and
Khochfar et al. simulations is not really a fundamental problem, as
both of these models do not include the impact of dust obscuration.

Although the Khochfar et al. simulation produces more star-formation rate density at higher redshift 
than either the Illustris or Munich models, the best predictions are produced by the fairly simple 
analytic models of Dayal et al. (2015), Cai et al. (2014) and Mason et
al. (2015). This might be regarded as unsurprising given that, as mentioned above,
these models are to some extent based on mappings between the dark
matter halo mass function and the UV luminosity function that have been `calibrated' 
at high redshift. Nevertheless, the fact that such models do such a good job in reproducing the rate of 
decline in $\rho_{UV}$ from $z \simeq 5$ to $z \simeq 10$ is surely not a coincidence, and implies 
that galaxy evolution at these early times is indeed fairly simple, and driven primarily by the 
growth in the underlying dark matter.

As final evidence in support of this basic result, in Fig.\,11 we compare our results with the 
generic analytic prediction of the high-redshift evolution of $\rho_{SFR}$ deduced by Hernquist \& Springel 
(2003). As a result of an effort to understand the physical processes driving the cosmic star formation
rate, Hernquist \& Springel (2003) predicted that at high redshift, beyond its peak, $\rho_{SFR}$ 
should decline smoothly in a manner very well approximated by the simple relation $\rho_{SFR} \propto
exp(z/3)$. In other words, in the absence of significant dust obscuration, they predicted 
a linear descent of $\log_{10}\rho_{UV}$ with $z$, and predicted the {\it slope} of this relation based on the 
assumption that the evolution is driven simply by the growth of the underlying dark matter.
In Fig.\,11 we show our derived evolution of $\rho_{UV}$ integrated down to $M_{1500}=-17.7$ 
(as previously plotted in Figs 9 and 10), but also show the results of integrating down by approximately 
another order-of-magnitude in luminosity, to $M_{1500}=-15$. Interestingly, while the results 
based on the shallower integration can, as already shown in Fig.\,9, be described by a linear relation between 
$\log_{10}\rho_{UV}$ and $z$, the slope of this relation is clearly steeper than predicted by 
Hernquist \& Springel (2003). However, by simply integrating down to $M_{1500}=-15$, to provide 
a more complete measurement of  $\rho_{UV}$ (especially at high redshift), it can be seen 
that the observed and predicted rate of decline of $\rho_{UV}$ shift into excellent agreement.

\section{Conclusions}
In this study we have assembled the largest sample of $z\sim 9-11$
galaxy candidates to date, identifying a total of 33 candidates in the
redshift range $8.4<z_{phot}<11.2$ from the HFF and CLASH
datasets. When combined with four candidates identified from our
previous analysis of the HUDF, our final sample consists of 37 galaxy candidates at
$z\geq 8.4$ selected from 29 independent WFC3/IR pointings (total area
$\simeq 130$ arcmin$^{2}$). Based on this sample we have produced an improved measurement of the
UV-selected galaxy LF at $z\simeq 9$ and placed initial constraints on
the LF at $z\simeq 10$. In addition, we have revisited the issue of the
decline in UV luminosity density at $z\geq 8$. Our main conclusions can be summarised as follows:

\begin{enumerate}
\item{Employing an analysis restricted to the uniformly deep, and low
    magnification, regions of the HFF+HUDF datasets,  we have derived an
    improved measurement of the UV-selected galaxy LF at $z\simeq
    9$. Based on this ``blank field'' method, our new results
    strengthen the evidence that the LF continues to evolve smoothly between $z=8$
    and $z=9$. Specifically, within this redshift interval, we find that the evolution of the LF can be equally
    well described by a 0.5 magnitude dimming in $M^{\star}$ (pure
    luminosity evolution) or a factor of $\simeq 2$ drop in
    $\phi^{\star}$ (pure density evolution).}

\item{As a consistency check, we have also used the full lensed area of our
    CLASH survey dataset to derive an independent estimate of the
    $z\simeq 9$ LF. The determinations of the $z\simeq 9$ LF based on
    the blank-field and lensed samples are found to be fully consistent.}

\item{Using the combined CLASH+HFF+HUDF datasets we have derived
    initial constraints on the UV-selected galaxy LF at $z\simeq
    10$. We find that the number density of $z\simeq 10$ galaxies at
    $M_{1500}\simeq -19.7$ is $\log_{10}\phi = -4.1^{+0.2}_{-0.3}$, a
    factor of $\simeq 2$ lower than at $z=9$.}

\item{Based on our new results we have revisited the issue of the
    decline in UV luminosity density at $z\geq 8$. We conclude that
    the data continue to support a smooth decline in $\rho_{UV}$ over
    the interval $6<z<10$, in agreement with simple models of galaxy
    evolution driven by the growth in the underlying dark matter halo
    mass function.}

\end{enumerate}

\begin{figure}
\centering
\includegraphics[width=\columnwidth, angle=0]{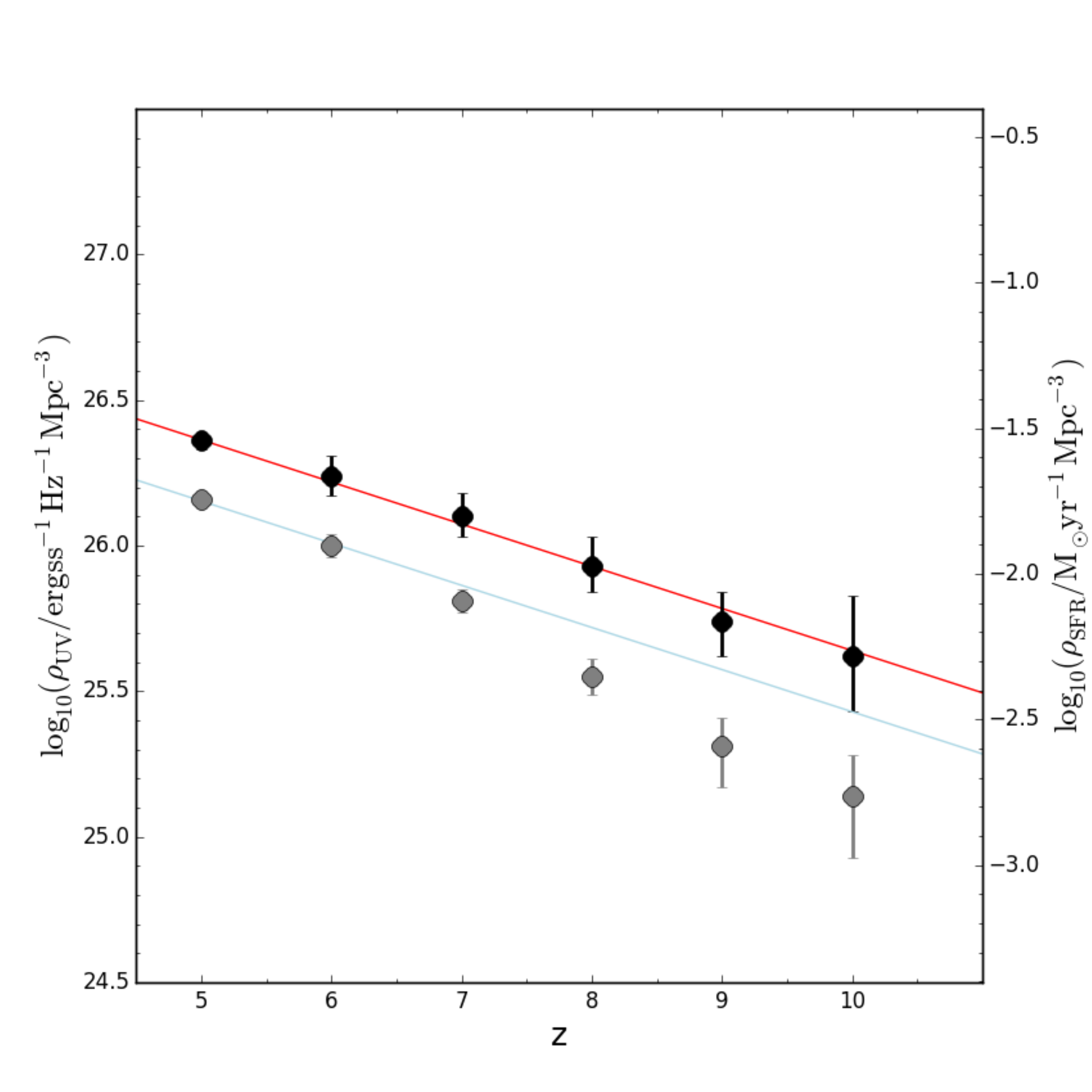}
\caption{Our measurement of the evolution of the comoving UV luminosity density at $z>5$ obtained by integrating  the luminosity-weighted LF down to a limit of $M_{1500}=-17.7$ (grey points, as already plotted in Figs. 9 \& 10) and by integrating 
down to a fainter luminosity limit of $M_{1500}=-15$ (black points).  The theoretical rate of decline of 
$\rho_{SFR} \propto\exp(z/3)$ predicted by the analytic work of Hernquist \& Springel (2003) is overplotted for both cases (scaled to match at $z \simeq 5$, to highlight the comparison between the observed 
and predicted rate of decline at higher redshift).  
While our results 
based on the shallower integration can, as already shown in Fig.\,9, be described by a linear relation between 
$\log_{10}\rho_{UV}$ and $z$, the slope of this relation is clearly steeper than predicted by 
Hernquist \& Springel (2003). However, by simply integrating down to $M_{1500}=-15$, to provide 
a more complete measurement of  $\rho_{UV}$ (especially at high redshift), it can be seen 
that the observed and predicted rate of decline of $\rho_{UV}$ shift into excellent agreement.}
\end{figure}

\section*{acknowledgements}
DJM and RJM acknowledge the support of the European Research Council via the award of a 
Consolidator Grant (PI McLure). JSD acknowledges the support of the European Research Council via the award of an Advanced Grant, 
and the contribution of the EC FP7 SPACE project ASTRODEEP (Ref. No.: 312725).
This work is based in part on observations made with the NASA/ESA {\it Hubble Space Telescope}, which is operated by the Association 
of Universities for Research in Astronomy, Inc, under NASA contract NAS5-26555.
This work is also based in part on observations made with the {\it Spitzer Space Telescope}, which is operated by the Jet Propulsion Laboratory, 
California Institute of Technology under NASA contract 1407.
\\\indent This work utilizes gravitational lensing models produced by PIs Brada\v c, Natarajan \& Kneib (CATS), Merten \& Zitrin, Sharon, and Williams funded as part of the HST Frontier Fields program conducted by STScI. STScI is operated by the Association of Universities for Research in Astronomy, Inc. under NASA contract NAS 5-26555. The lens models were obtained from the Mikulski Archive for Space Telescopes (MAST).  For CLASH we also used mass models constructed by A. Zitrin et al. (2013, 2015),
obtained through the Hubble Space Telescope Archive, as a high-end science product.

{}

\end{document}